\documentclass[aps,pre,twocolumn,showpacs,superscriptaddress]{revtex4-1}  
\usepackage{graphicx}
\usepackage{dcolumn}
\usepackage{bm}   
\usepackage{color}
\usepackage{amssymb}
\usepackage{amsmath}

\newcommand{\sech}{\normalfont\mbox{sech}\,}

\begin{document}

\title{Manipulating matter-rogue waves and breathers in Bose-Einstein condensates}
\author{K. Manikandan}
\affiliation{Centre for Nonlinear Dynamics, Bharathidasan University, Tiruchirappalli 620024, Tamilnadu, India}
\author{P. Muruganandam}
\affiliation{Department of Physics, Bharathidasan University, Tiruchirappalli 620024, Tamilnadu, India}
\author{M. Senthilvelan}
\affiliation{Centre for Nonlinear Dynamics, Bharathidasan University, Tiruchirappalli 620024, Tamilnadu, India}
\author{M. Lakshmanan}
\affiliation{Centre for Nonlinear Dynamics, Bharathidasan University, Tiruchirappalli 620024, Tamilnadu, India}

\begin{abstract}
We construct higher order rogue wave solutions and breather profiles for the quasi-one-dimensional Gross-Pitaevskii equation with a time-dependent interatomic interaction and external trap through the similarity transformation technique.  We consider three different forms of traps, namely (i) time-independent expulsive trap, (ii) time-dependent monotonous trap and (iii) time-dependent periodic trap.  Our results show that when we change a parameter appearing in the time-independent or time-dependent trap the second and third-order rogue waves transform into the first-order like rogue waves.  We also analyze the density profiles of breather solutions.  Here also we show that the shapes of the breathers change when we tune the strength of trap parameter.  Our results may help to manage rogue waves experimentally in a BEC system.
\end{abstract}
\pacs{03.75.Kk, 03.75.Lm, 67.85.Hj, 05.45.Yv}
\maketitle

\section{Introduction}
During the past several years considerable interest has been shown in exploring localized nonlinear waves in the variable coefficient nonlinear Schr\"{o}dinger (NLS) equation and its generalizations \cite{pono,serki,cent}.  The motivation comes from the fact that NLS equation and its variants appear in several branches/topics of physics, including nonlinear optics \cite{hase,solli} and Bose-Einstein condensates (BECs) \cite{peth:smi,dalf,carr}, etc.  Focusing our attention on BECs alone, it is well known that the Gross-Pitaevskii (GP) equation governs the evolution of macroscopic wave function at ultra low temperatures \cite{peth:smi,pit:str}.  In particular, for cigar shaped BECs, it has been shown that the GP equation can be reduced to the 1D variable coefficient NLS equation \cite{perez, raj:mur, atre:pani}
\begin{align}
i\psi_t+\frac{1}{2}\psi_{xx}+R(t)\vert \psi\vert ^2\psi+\frac{1}{2} \beta(t)^2 x^2 \psi=0, \label{eq:1d-gp}
\end{align}
where $\psi(x,t)$ is the condensate wave function, $t$ is the dimensionless normalized time, $x$ is the dimensionless normalized coordinate in the axial direction, $R(t)$ represents the effective scattering length and $\beta(t)$ is the axial trap frequency.

A simple and straightforward way of exploring the localized/periodic structures of (\ref{eq:1d-gp}) is by transforming it into a constant coefficient NLS equation through a suitable transformation.  From the known solutions of the latter equation the solutions of the former equation can be identified.  Using this procedure a class of solutions, in particular various localized solutions, have been identified for the model (\ref{eq:1d-gp}), including soliton, breather, and rogue wave (RW) solutions.  In order to appreciate the relevance of the above type of localized structures for Eq. (\ref{eq:1d-gp}), we may first consider their existence in the case of constant coefficient NLS equation, 
\begin{align}
iU_{T}+\frac{1}{2}U_{XX}+ \vert U\vert ^2U=0. \label{nls}
\end{align}
It is well known that the standard NLS equation (\ref{nls}) admits the following basic localized profiles and their higher-order versions \cite{mlsr}. 

\begin{figure*}[!ht]
\begin{center}
\includegraphics[width=0.8\linewidth]{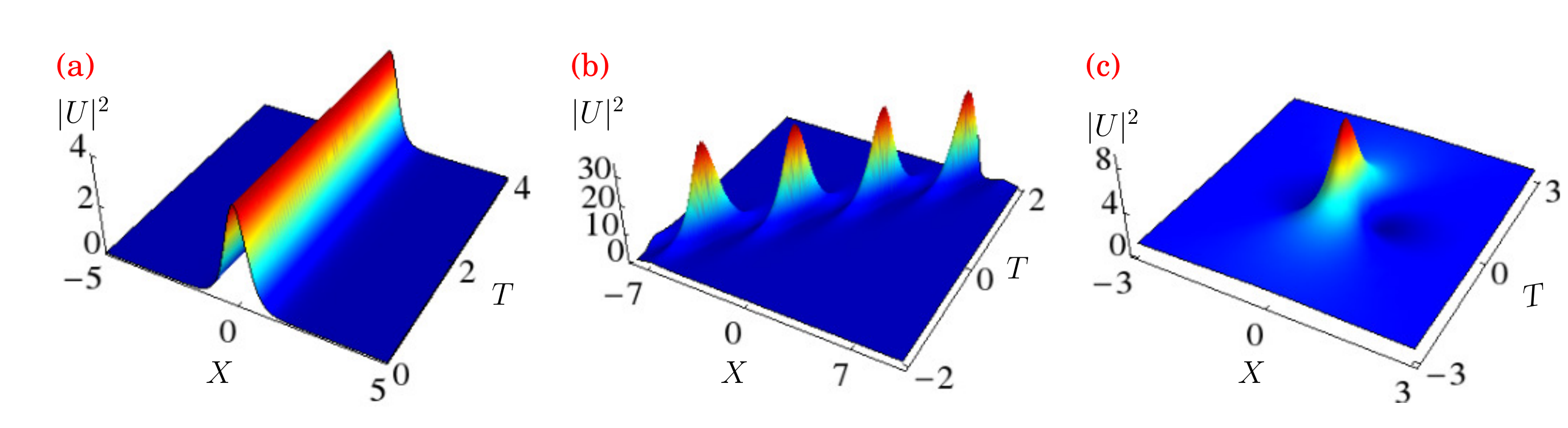}
\end{center}
\caption{(Color online) Profiles of (a) Soliton, (b) General breather, and (c) Peregrine soliton (rogue wave).}
\label{figa}
\end{figure*}

\paragraph*{(i) Envelope soliton:} It is a solitary wave (localized envelope along with a carrier wave) that retains its characteristics (amplitude/shape and velocity) under collision with another soliton, except for a change in phase.  The intensity profile of the soliton is shown in Fig. \ref{figa}(a).  The typical form of the envelope soliton is 
\begin{align}
U(X,T)= P_{1R} \exp^{i \eta_1} \sech{\eta_2}, \label{nls1}
\end{align}
where $P_1(=P_{1R}+iP_{1I})$ is a complex constant, $\eta_1=P_{1I} X+[(P_{1R}^2-P_{1I}^2)T]/2+\eta_{1I}^{(0)}$ and $\eta_2=P_{1R}(X- P_{1I} T)+\eta_{1R}^{(0)}+\log{(1/2P_{1R})}$.  Here $\eta_{1R}^{(0)}$ and $\eta_{1I}^{(0)}$ are constant parameters.
   
\paragraph*{(ii) Breather:} It is a localized solution with temporally and/or spatially periodic structures having constant background exhibiting internal oscillations and bound states of nonlinear wave packets \cite{mand} which is represented in Fig. \ref{figa}(b). Its typical form reads \cite{tajri} 
\begin{widetext}
\begin{align}
U(X,T) &= \rho_0 \cos{(2\phi_R)}\exp{i(\theta+2\phi_R)} \left[1+\frac{1}{\sqrt{a}\cosh{(\eta_R+\sigma)}+\cos{\eta_I}} \right. \notag \\ &   \left. \times\left\{\left(\frac{\cosh{2\phi_I}}{\cos{2\phi_R}}-1\right)\cos{\eta_I}+i\left(\tan{2\phi_R}\sinh{(\eta_R+\sigma)}-\frac{\sinh{2\phi_I}}{\cos{2\phi_R}}\sin{\eta_I}\right)\right\}\right] 
\label{nls2}
\end{align}
\end{widetext}
where $\rho_0$ is the amplitude of the plane wave, $\theta=kX-\omega T$, $\sigma$ , $k$, $\omega$ are constant parameters, $\eta_R=P_R X-(\Omega_R T)/2+\eta_R^0$, $\eta_I=P_I X-(\Omega_I T)/2+\eta_I^0$, $\Omega_R=2kP_R-\frac{P_R^2-P_I^2\sin{2\phi_R}+2P_RP_I\sinh{2\phi_I}}{\cosh{2\phi_I}-\cos{2\phi_R}}$, $\Omega_I=2kP_I+\frac{P_R^2-P_I^2\sin{2\phi_I}+2P_RP_I\sinh{2\phi_R}}{\cosh{2\phi_I}-\cos{2\phi_R}}$, $a=\cosh^2{\phi_I}/\cos^2{\phi_R}$, $P_R=-2\rho_0\cos{\phi_R}\sinh{\phi_I}$, $P_I=2\rho_0\sin{\phi_R}\cosh{\phi_I}$, and $\phi=\phi_R+i\phi_I$ is a complex constant. This solution is also called a general breather (GB) since it is periodic both in space and in time. Two important special cases are the following:  (i) When $\phi_R\neq 0$ and $\phi_I=0$, the GB solution corresponds to an Akhmediev breather (AB) which is periodic in space and localized in time.  (ii) If we take $\phi_R=0$ and $\phi_I\neq 0$, (\ref{nls2}) becomes a Ma breather (MB) which is periodic in time only and localized in space.
\paragraph*{(iii) Rogue wave:} A further specialized structure, which is localized both in space and in time with a constant plane wave background, is the so called Peregrine soliton or RW \cite{pere}.  It can be obtained by taking the limits $\phi_R=\epsilon\gamma$ and $\phi_I=\epsilon\delta$, and $\epsilon\to 0$, where $\gamma$, $\delta$ and $\epsilon$ are constants.  Then Eq. (\ref{nls2}) reduces to the following form,
\begin{align}
U(X,T)=\rho_0 \,\text{e}^{ i \theta }\left(1-\frac{4+8i\rho_0^2 T}{1+4\rho_0^2(X-kT)^2+4\rho_0^4T^2}\right). \label{nls3}
\end{align}
Very often the above localized nonlinear wave is described as a wave that ``appears from nowhere and disappears without a trace" \cite{osbrn:rato} as shown in Fig. \ref{figa}(c). It was first observed in the area of oceanography \cite{osbrn} and is traditionally defined as a wave whose wave-height (the distance from trough to crest) is more than twice the significant wave height (SWH).  The latter is generally defined as the average wave height among one third of highest waves in a given time series \cite{osbrn,khar:pelin}.  Further, it has also been explained that the above structure (\ref{nls3}) arises due to a modulation instability (MI) \cite{pere,benj:feir} of the plane wave solution of the constant coefficient NLS equation.  Very recently higher-order RWs (HRWs) which correspond to the higher-order rational solutions of the NLS equation (\ref{nls}) have been deduced \cite{akmv:anki}.  The explicit expressions for the second and third-order RWs of the NLS equation are given in appendix A.  These HRWs have higher amplitudes than the first-order RW.  The RWs have also been observed experimentally in physical systems such as water wave tank \cite{chab}, capillary waves \cite{shatz} and nonlinear optics \cite{solli,kibler}.  Several theoretical studies on the dynamics of RWs in nonlinear fiber optics \cite{porse,pors}, plasma physics \cite{mose}, laser-plasma interactions \cite{veldas}, and even econophysics \cite {tan}, described by scalar NLS equation, have been made in recent times.

Now, as mentioned in the beginning, the dynamics of a cigar shaped BEC at absolute zero temperature is usually described by the mean-field GP equation (\ref{eq:1d-gp}), which is a generalized form of the ubiquitous constant coefficient NLS equation (\ref{nls}), for the wave function of the condensate.  Since the NLS equation (\ref{nls}) admits breather and RW solutions, it is natural to expect that RWs and breathers may also be found in BEC systems as well. In this context, the RWs can correspond to a sudden increase of peaks in the condensate clouds similar to the nature of high peaks in the open sea, while breathers are generalizations of the RWs.  The formative mechanism for the matter RWs in BECs is the accumulation of energy and atoms towards its central part and their spreading out to a constant density background.  The formation of matter breather is the periodic exchange of atoms between the profile and the plane wave background.  From an experimental point of view, the existence of RW and breather structures in a BEC system can be effectively controlled by tuning the nonlinear interaction between atoms by Feshbach resonance technique \cite{dalf,stali,abdul} and modulating the trapping frequency of the external potential.  It will therefore be of great interest to study the characteristics and/or controlling of structures of RWs and breathers due to their localization both in space and in time in BEC experiments. Past explorations of GP equation in BECs have paved the way for important developments in manipulating coherent matter waves for application, including atom interferometry \cite{inter}, coherent atom transport \cite{mandel} and quantum information processing or quantum computation \cite{cirac}.  Therefore, it is of high significance to study the dynamics of RWs and breather profiles of the GP equation (\ref{eq:1d-gp}).  However only few attempts have been made to identify and analyze the RWs and breather solutions of (\ref{eq:1d-gp}) \cite{blud:kono,yan,wen:li,zhao, wu,he}.  To the best of our knowledge neither higher-order RW solutions (with certain free parameters) nor higher order breather solutions of (\ref{eq:1d-gp}) have been taken up for study.  Motivated by these observations, in this work, we construct the aforementioned localized and periodic solutions of (\ref{eq:1d-gp}).  Besides constructing these two families of solutions we also investigate how to manipulate the RWs and breathers through the effective scattering length and the strength of trap parameter.  

Having stated the motivation we now proceed to construct a transformation that transforms Eq. (\ref{eq:1d-gp}) to the standard NLS equation (\ref{nls}).  Following the standard procedure \cite{raj:mur,raj:ml} we find that the required similarity transformation should be of the form
\begin{align}
\label{a7}
\psi(x,t)&  = r_0\sqrt{R(t)}[U(X,T)] \notag \\ 
& \times \exp \left[{i\left(c_1r_0^2 Rx-\frac{R_t}{R}x^2-\frac{1}{2}c_1^2r_0^4\int{R^2(t)}dt\right)}\right],
\end{align}
where 
\begin{subequations}
\label{a4}
\begin{align}
X(x,t)= & \, r_0 R(t)x-c_1r_0^3\int{R^2(t)}dt,  \\
T(t)= & \, r_0^2\int{R^2(t)}dt,
\end{align}
\end{subequations}%
and $U(X,T)$ is the solution of the standard NLS equation (\ref{nls}).  In the above $b,r_0$ are arbitrary constants and the modulational functions $R(t)$ and $\beta(t)^2$ should satisfy the following condition
\begin{align}
\label{a5}
\frac{d}{dt}\left(\frac{R_t}{R}\right)-\left(\frac{R_t}{R}\right)^2+\beta(t)^2=0,
\end{align}
which is a Riccati type equation with dependent variable $(R_t/R)$ and independent variable $t$.  Regardless of what $R(t)$ is, as long as the condition (\ref{a5}) is satisfied, the GP equation is integrable \cite{raj:mur, raj:ml}.  We also note here that the Painlev\'e singularity structure analysis performed on Eq. (\ref{eq:1d-gp}) confirms the same restriction (\ref{a5}) on the system coefficients \cite{pain}.  We further note that the solution (\ref{a7}) provides us some flexibility to generate new structures related to the RWs which may be useful for the BEC experiments. 

Even though one can arbitrarily choose the functions $R(t)$ and $\beta(t)$ that satisfy the constraint (\ref{a5}) and generate the required solutions, in this paper we consider the trap frequency to be of the following three forms: (i) $\beta(t)^2=\beta_0^2$, (ii) $\beta^2(t)=\left({\beta_0^2}/{2} \right)\left[1-\tanh\left({\beta_0 t/}{2}\right)\right]$ and (iii) $\beta(t)^2=2\beta_0^2[1+3\tan^2(\beta_0 t)]$, where $\beta_0$ is a constant.  As shown in reference \cite{raj:mur} the effective scattering length for these three cases turn out to be (i) $R(t)=\sech{(\beta_0 t+\delta)}$, (ii) $R(t)=1+\tanh \left({\beta_0 t}/{2}\right)$ and (iii) $R(t)=1+\cos{(2 \beta_0 t)}$.  We consider the trap frequency to be in the above forms since it has been shown that they are valid forms in BEC experiments \cite{raj:mur}.  We consider each one of the cases separately and substitute them in (\ref{a7}) along with the RW and breather solutions of (\ref{nls}).  We then analyze in detail how the nature of the RW and breather structures get modified by the above two functions $\beta(t)$ and $R(t)$.  Our analysis shows that the amplitude parameter $(r_0)$ plays a vital role in the formation of RW, and the trap frequency and the effective scattering length modify the structure of the RW and breather profiles, allowing one the possibility of manipulating the RWs and breathers in specific ways. 

The paper is organized as follows.  In the following section, we construct RW solutions for time-independent and time-dependent traps and study their characteristics in detail.  To the best of our knowledge, for the first time in the literature, we observe that the second- and third-order RWs transform to first-order RW-like structures when we tune a parameter which appears in the harmonic traps (time-independent and time-dependent traps).  In Sec. III, we also construct RW solutions with free parameters which allow us to split the symmetric form solution into a multi-peaked solution for (\ref{eq:1d-gp}) and investigate how these RW structures get modified in the plane wave background by increasing the strength of the trap.  In Sec. IV, we construct one-breather and two-breather solutions of (\ref{eq:1d-gp}) and investigate their characteristics when we alter the trap parameter.  Finally, in Sec. V, we present a summary of the results and our conclusions. 
\section{Characteristics of rogue waves}
To begin with, we consider the case in which the trap frequency is a constant, that is $\beta(t)^2$ = constant = $\beta_0^2$.  A time-independent trap frequency implies that the frequency does not change with time and space. We then consider the trap frequency to be time-dependent and investigate the associated RW solutions.  
\subsection{Time-independent trap} 
Substituting $\beta(t)^2=\beta_0^2$ in the integrability condition $(\ref{a5})$, we find that the time-dependent interaction term should be of the form $R(t)=\sech{(\beta_0 t+\delta)}$, where $\delta$ is an integration constant.   
\begin{figure*}[!ht]
\begin{center}
\includegraphics[width=0.8\linewidth]{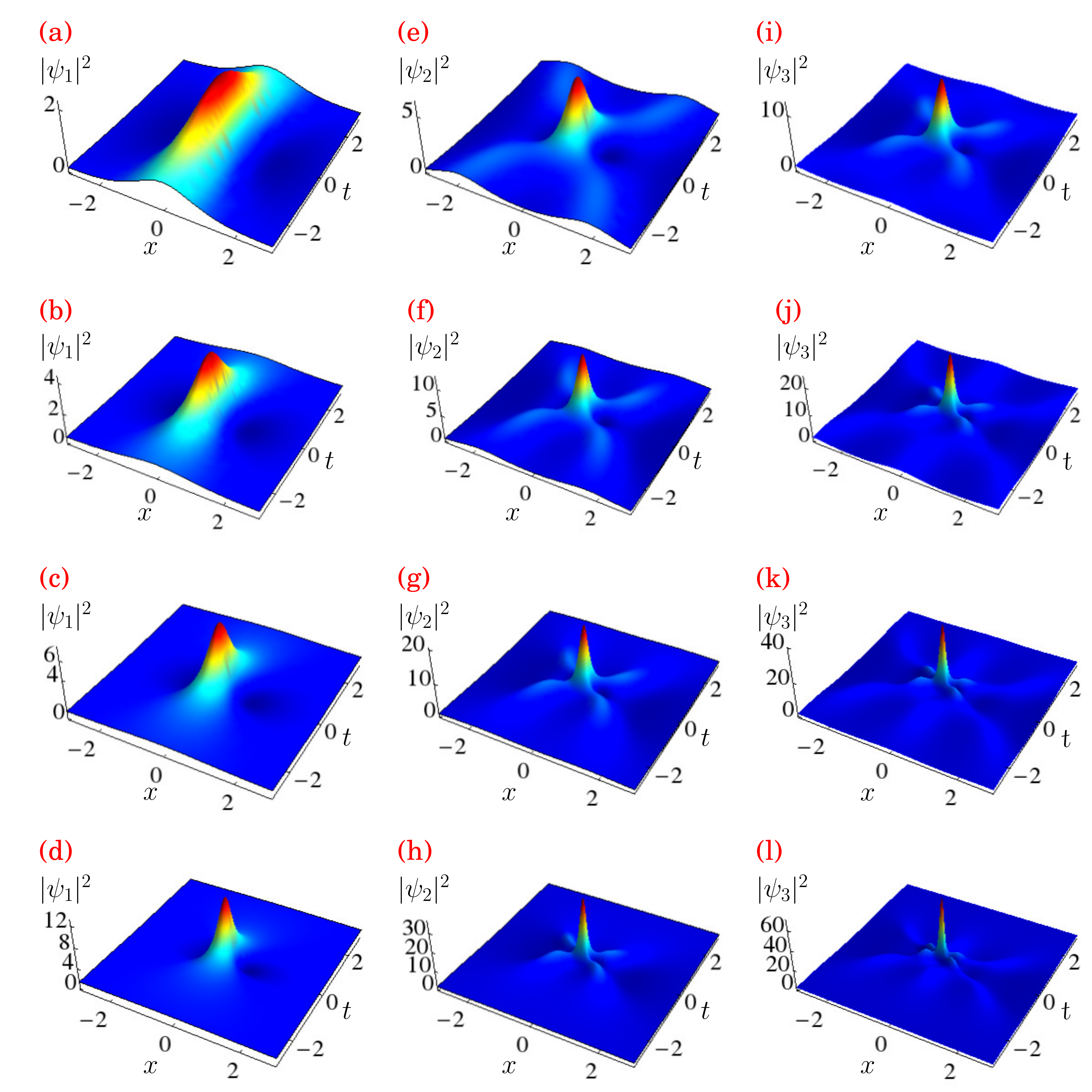}
\end{center}
\caption{(Color online) First, second, and third columns represent the formation of first-, second-, and third-order RWs in BECs for the time-dependent nonlinearity coefficient $R(t)=\sech{(\beta_0 t+\delta)}$ and time-independent trap frequency $\beta(t)^2=\beta_0^2$, obtained using the expression (\ref{a15}).  The parameter $r_0=0.5$ for Figs. (a), (e) and (i), $0.7$ for (b), (f) and (j), $0.9$ for (c), (g) and (k), and 1.2 for (d), (h) and (l). The other parameters are $c_1=0.01$, $\beta_0=0.1$, and $\delta=0.01$.}
\label{fig8}
\end{figure*}
\begin{figure}[!ht]
\begin{center}
\includegraphics[width=\linewidth]{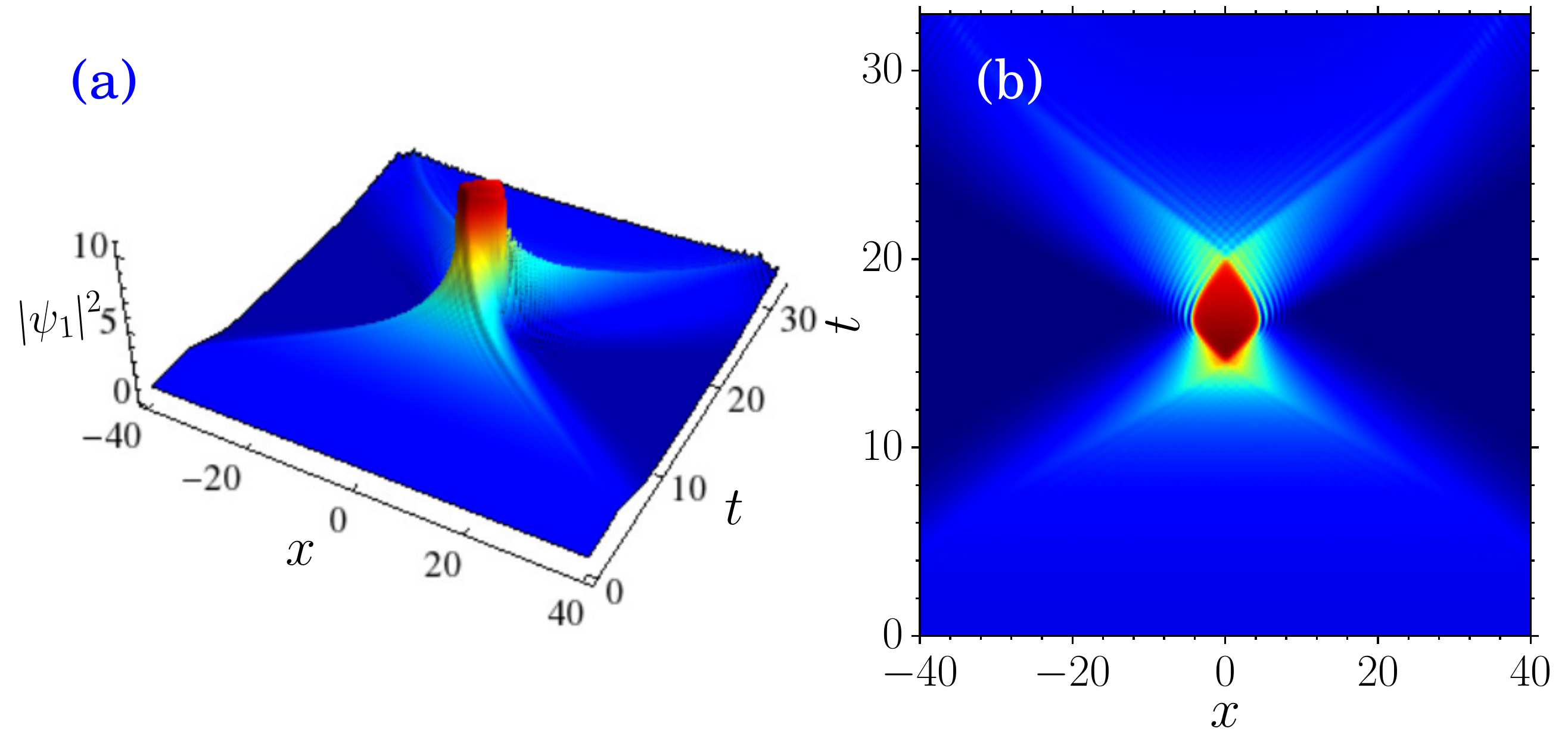}
\end{center}
\caption{(Color online) (a) First-order RW in BEC and (b) the corresponding contour plot obtained by numerically solving Eq. (\ref{eq:1d-gp}) through split-step Crank-Nicolson method for the time-dependent nonlinearity coefficient $R(t)=\sech{(\beta_0 t+\delta)}$ and time-independent trap frequency $\beta(t)^2=\beta_0^2$. The initial condition chosen corresponds to the analytic solution of Fig.~\ref{fig8}(d).}
\label{num-rog}
\end{figure}
Plugging this expression in (\ref{a7}), we find
\begin{align}
\label{a15}
\psi_j(x,t) = r_0\sqrt{\sech{(\beta_0 t+\delta)}}\, U_j(X,T)  \eta(x,t),
\end{align}
where
\begin{align}
\eta(x,t) = & \exp\bigg\{i\bigg[c_1 r_0^2 \sech{(\beta_0 t+\delta)}x  \notag \\
& \left.\left. +\frac{(\beta_0^2 x^2-c_1^2r_0^4)\tanh{(\beta_0 t+\delta)}}{2\beta_0}\right]\right\}, \notag
\end{align}
and $U_j(X,T)$'s, $j=1, 2, 3$, are the first, second and third-order RW solutions of the NLS equation (\ref{nls}) whose explicit expressions are given in the Appendix (vide Eqs. (\ref{a8}), (\ref{a11}) and (\ref{a12})). Also $X(x,t)$ and $T(t)$ have the forms as given in Eqs. (\ref{a4}).
\begin{figure*}[!ht]
\begin{center}
\includegraphics[width=0.8\linewidth]{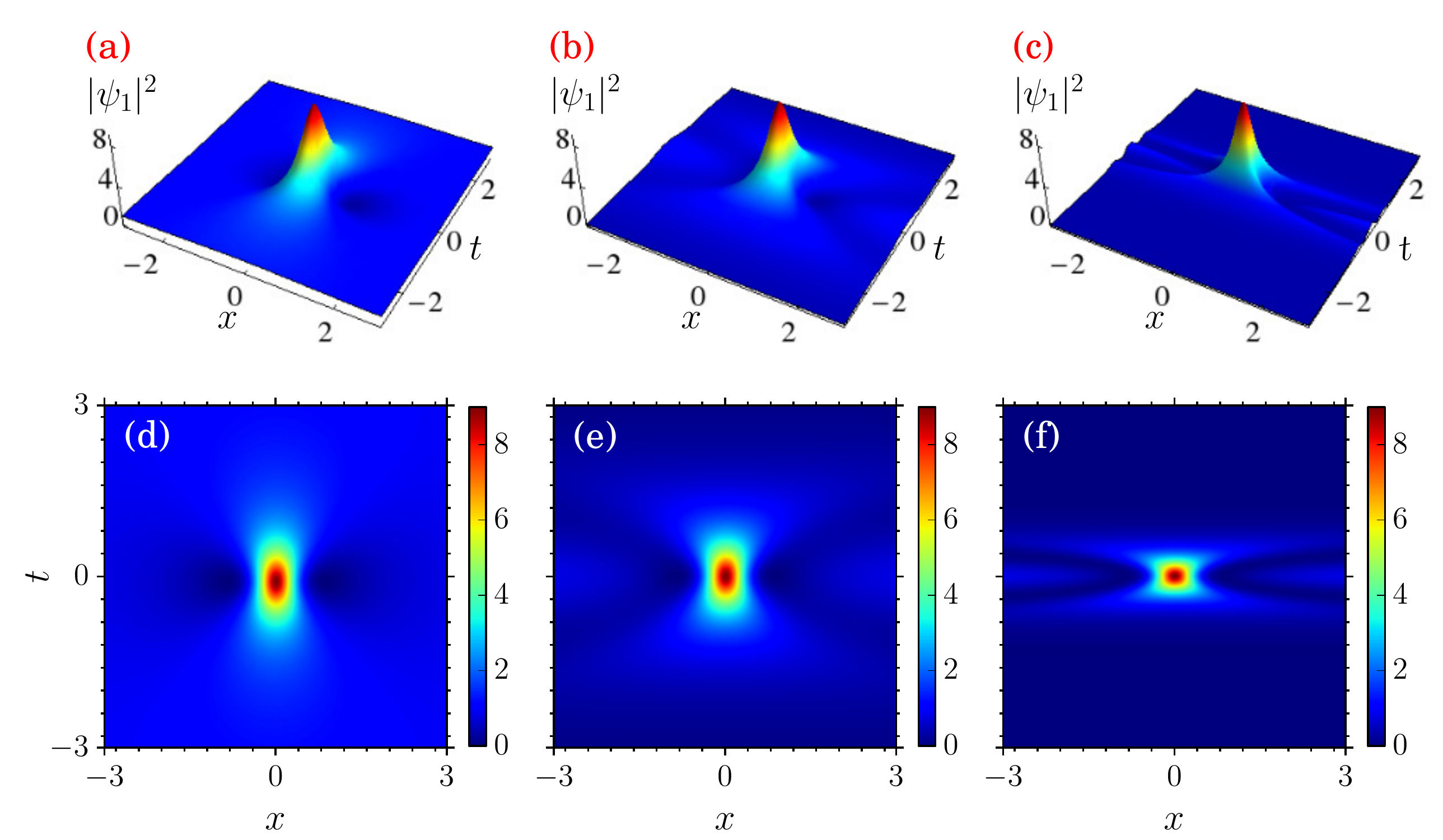}
\end{center}
\caption{(Color online) First-order RWs for $R(t)=\sech{(\beta_0 t+\delta)}$ and $\beta(t)^2=\beta_0^2$.  The parameter $\beta_0$ is varied as (a) $\beta_0=0.1$, (b) $\beta_0=1.2$, (c) $\beta_0=5.0$.  Figs. (d), (e) and (f) are their corresponding contour plots. The other parameters are fixed as $r_0=1.0$, $c_1=0.01$, and $\delta=0.01$.}
\label{fig1}
\end{figure*}
\begin{figure*}[!ht]
\begin{center}
\includegraphics[width=0.8\linewidth]{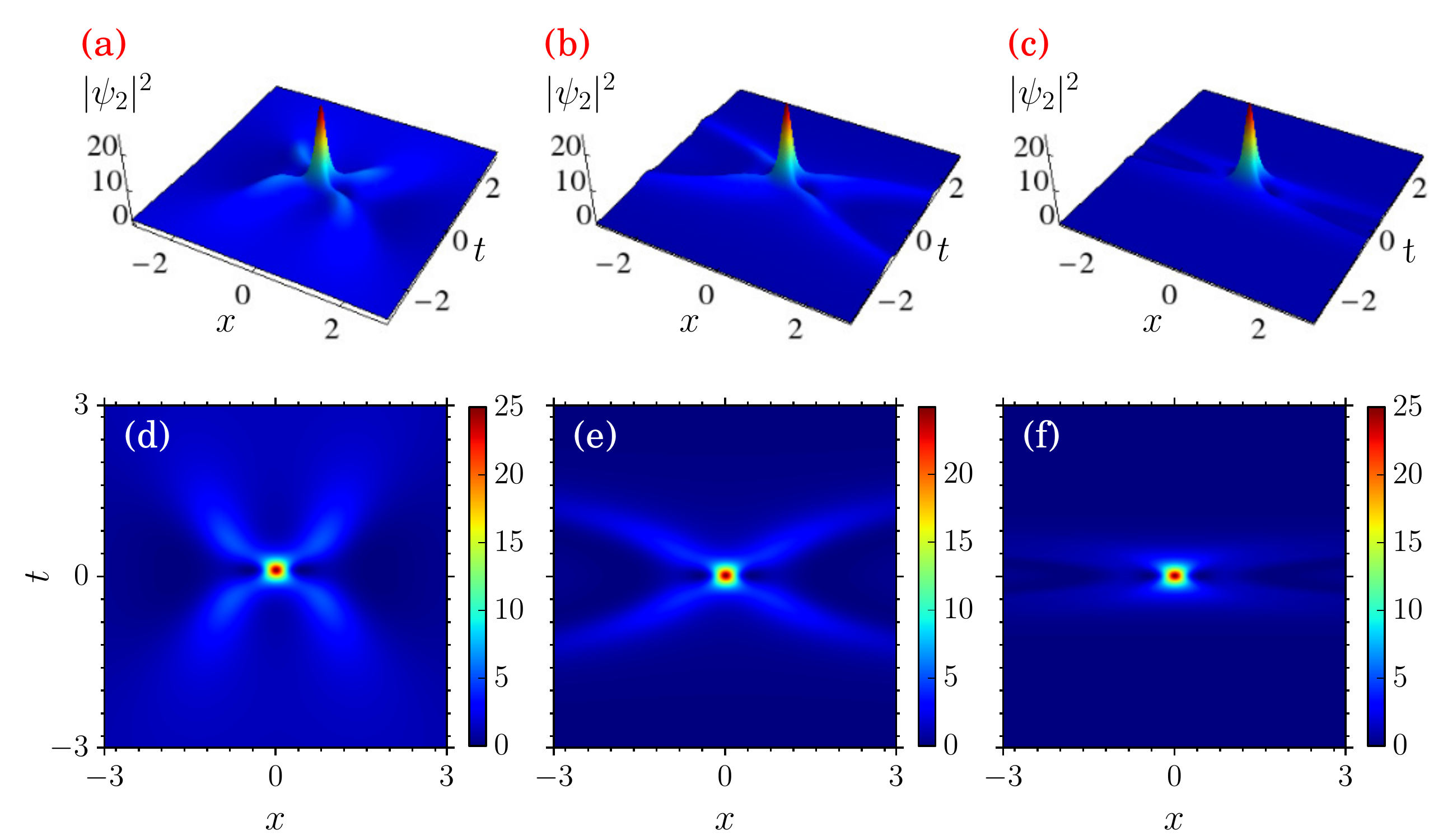}
\end{center}
\caption{(Color online) Second-order RWs for $R(t)=\sech{(\beta_0 t+\delta)}$ and $\beta(t)^2=\beta_0^2$.  The parameter $\beta_0$ is varied as (a)$\beta_0=0.1$, (b) $\beta_0=1.2$, (c) $\beta_0=5.0$.  Figs. (d), (e) and (f) are their corresponding contour plots. The other parameters are same as in Fig.~\ref{fig1}.}
\label{fig1a}
\end{figure*}

\begin{figure*}[!ht]
\begin{center}
\includegraphics[width=0.8\linewidth]{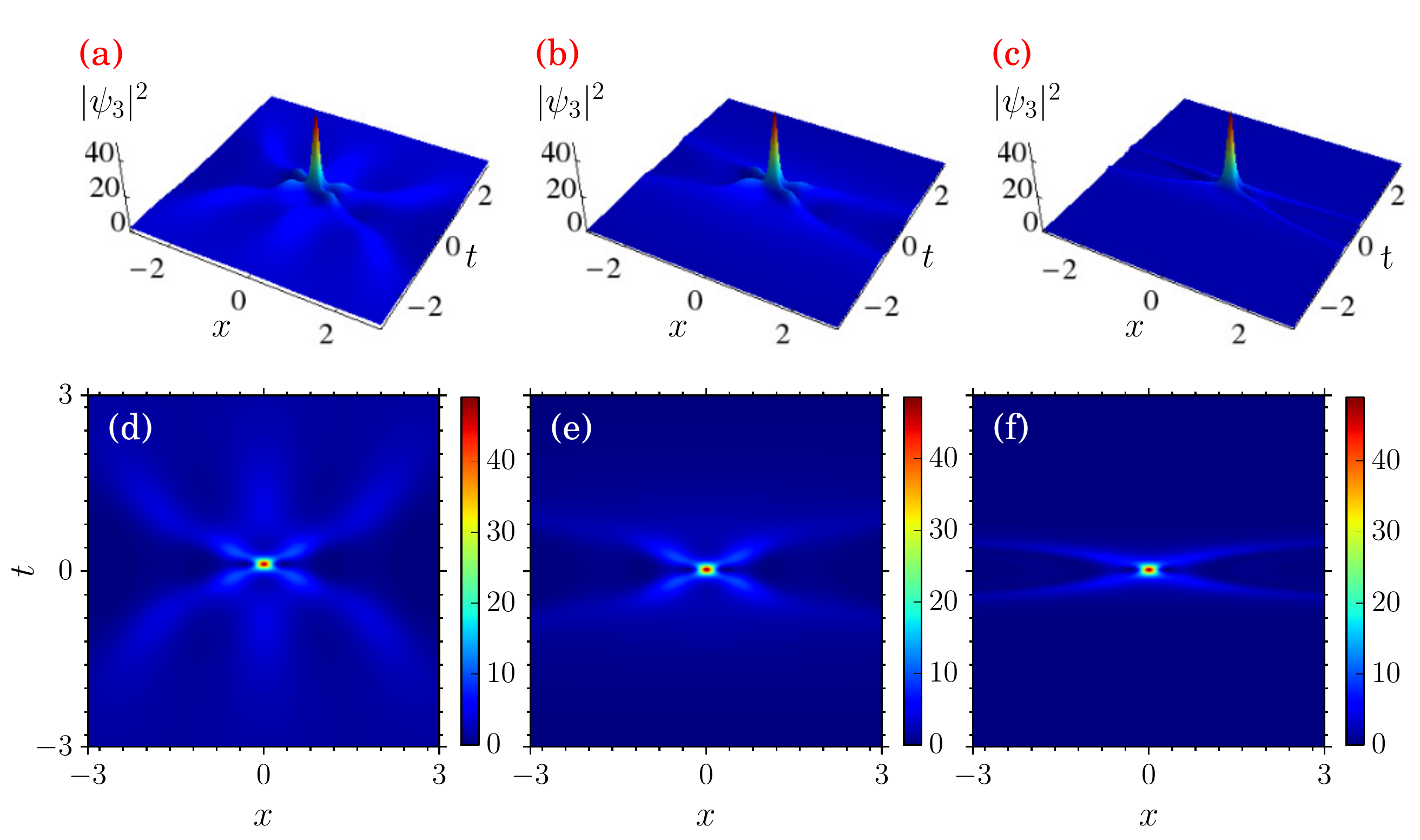}
\end{center}
\caption{(Color online) Third-order RWs for $R(t)=\sech{(\beta_0 t+\delta)}$ and $\beta(t)^2=\beta_0^2$.  The parameter $\beta_0$ is varied as (a) $\beta_0=0.1$, (b) $\beta_0=1.5$, (c) $\beta_0=5.0$.  Figs. (d), (e), and (f) are their corresponding contour plots. The other parameters are same as in Fig.~\ref{fig1}.}
\label{fig3a}
\end{figure*}

In Fig. \ref{fig8}, the first, second, and third columns represent the density profiles of the first, second, and third-order RWs obtained from (\ref{a15}). In this figure we present the formation of RWs in cigar-shaped BECs.  In the context of BECs, it is the fluctuation in the density of atoms, localized in space and time, which is what we observe as RWs.  For instance consider the first-order RWs depicted in Figs. \ref{fig8}(a)-(d). These may be interpreted as follows: Atoms in the condensate, initially at a constant density background, suddenly accumulate to form a hump towards the center of the condensate at finite time while leaving voids in the density which appear as troughs in the RWs, depending on the initial state. The crest and troughs reach their extrema as time progresses, and then the condensate atoms spread out so as to recover the constant density background in finite time, thus revealing the unstable nature of RWs. Further, by tuning the amplitude parameter $r_0$ in (\ref{a15}), we can visualize the formation and manipulation of RWs.  For example, when we increase the amplitude parameter $r_0$ smoothly from $0.5$ to $1.2$, we can observe in Figs.~\ref{fig8}(a)-(d) the formation of more and more localized first-order RWs with increasing amplitude. At $r_0=1.2$ one can visualize a large amplitude wave which is sufficiently localized both in space and in time which in turn confirms the formation of the first-order RW in BEC.  The formation of second and third-order RWs is also demonstrated in Figs. \ref{fig8}(c)-(h) and \ref{fig8}(i)-(l), respectively, for the same set of parameter values of $r_0$.  From these figures we infer that the time-dependent nonlinear interaction between the atoms induces density fluctuations over the condensate which gets more and more localized both in space and in time as we increase the order of the RW.

In order to confirm the existence of the RWs further, we have also performed a direct numerical simulation of (\ref{eq:1d-gp}) with the aid of the split-step Crank-Nicolson method using an initial wave function which is the same as the function (\ref{a15}) and with space step $dx=0.01$ and time step $dt=0.001$~\cite{Muruganandam+}.  In Fig.~\ref{num-rog}, we have presented the computer generated density profile of the first-order RW and the corresponding contour plot with the parameters chosen as $r_0=1.2$, $c_1=0.01$, $\beta_0=0.1$, and $\delta=0.01$ which are same as that of Fig.~\ref{fig8}(d). One can easily observe a very good agreement between the numerical results and the analytical predictions for the emergence of the RWs.  We have also verified numerically the presence of second- and third-order RWs of (\ref{eq:1d-gp}) as well, replicating Figs.~\ref{fig8}.

Next we demonstrate how these localized structures vary with respect to the trap parameter $\beta_0$.  Fig.~\ref{fig1} displays the first-order RW for the same nonlinearity management parameter $R(t)=\sech{(\beta_0 t+\delta)}$ and the external trap frequency $\beta(t)^2=\beta_0^2$. The nature of the first-order RW for $\beta_0=0.1$ is depicted in Fig.~\ref{fig1}(a).  When we increase the strength of the trap parameter the density profiles corresponding to the first-order RW become more and more localized in time as shown in Figs. \ref{fig1}(b) and \ref{fig1}(c), respectively. The corresponding contour plots are given in Figs.~\ref{fig1}(d)-(f).

Fig.~\ref{fig1a} displays the density profiles of the second-order RWs for the same nonlinearity management parameter as a function of $\beta_0$. In Fig.~\ref{fig1a}(a) we display the second-order RW for $\beta_0=0.1$.  When the strength of the parameter $\beta_0$ is increased to $1.2$, the wave subcrests start to stretch as shown in Fig.~\ref{fig1a}(b).  The wave subcrests become more and more localized in time when we increase the value of $\beta_0$ and finally the second-order RW attains a new structure as given in Fig.~\ref{fig1a}(c) for $\beta_0 = 5.0$.  The resultant structure looks almost like a first-order RW, see Fig.~\ref{fig1}(c).  Figs.~\ref{fig1a}(d)-(f) are the corresponding contour plots.

We also observe similar effects in the third-order RW case as well when we increase the value of $\beta_0$.  Fig.~\ref{fig3a} demonstrates the changes in the third-order RW when we vary the interaction strength.  At $\beta_0=1.5$, the third-order RW transforms into the second-order RW-like structure as shown in Fig.~\ref{fig3a}(b).  When we increase the value of the parameter $\beta_0$ to $5.0$, one obtains a first-order like RW which is displayed in Fig.~\ref{fig3a}(c). These facts are also confirmed by the corresponding contour plots, Figs.~\ref{fig3a}(d)-(f).  The aforementioned results reveal that when we increase the strength of the trap parameter $\beta_0$, the second and third-order RWs become more localized in time and delocalized in space, approaching the structure of a first-order RW. Thus the robustness of the density profiles can be controlled by varying the strength of trap frequencies. 
 
\subsection{Time-dependent monotonous trap}

Next we consider the time-dependent trap frequency in the form $\beta^2(t)=\left({\beta_0^2}/{2} \right)\left[1-\tanh\left({\beta_0 t/}{2}\right)\right]$. 
\begin{figure*}[!ht]
\begin{center}
\includegraphics[width=0.8\linewidth]{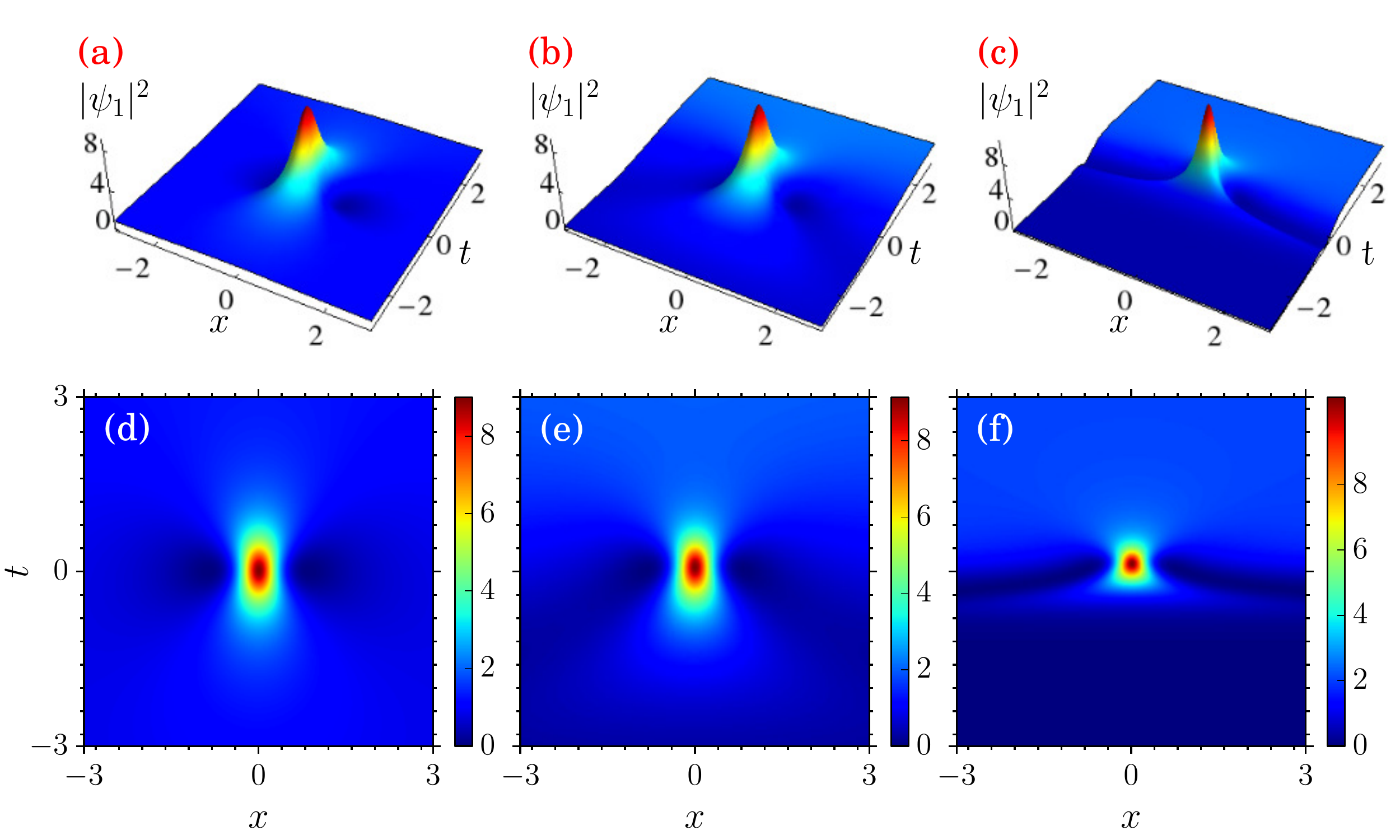}
\end{center}
\caption{(Color online) First-order RWs for $R(t)=1+\tanh\big({\beta_0 t}/{2}\big)$ and $\beta(t)^2={\beta_0^2}/{2}\left[1-\tanh\left({\beta_0 t}/{2}\right)\right]$. The parameter $\beta_0$ is varied as (a)$\beta_0=0.1$, (b) $\beta_0=1.0$ and (c) $\beta_0=5.0$.  Figs. (d), (e), and (f) are the corresponding contour plots of (a), (b) and (c). The other parameters are same as in Fig.~\ref{fig1}.}
\label{fig4}
\end{figure*}
\begin{figure*}[!ht]
\begin{center}
\includegraphics[width=0.9\linewidth]{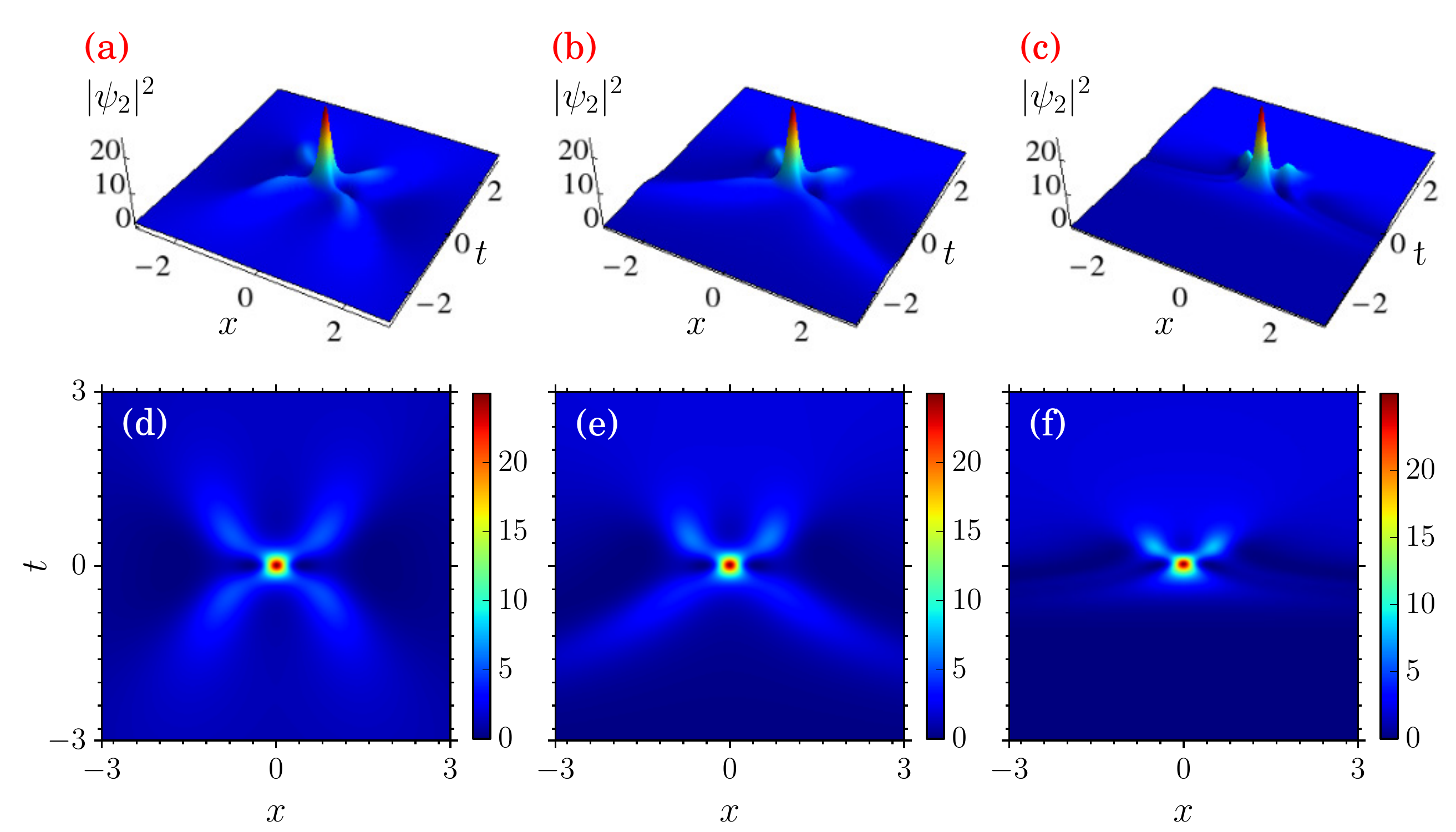}
\end{center}
\caption{(Color online) Second-order RWs for $R(t)=1+\tanh\big({\beta_0 t}/{2}\big)$ and $\beta(t)^2={\beta_0^2}/{2}\left[1-\tanh\left({\beta_0 t}/{2}\right)\right]$. The parameter $\beta_0$ is varied as (a)$\beta_0=0.1$, (b) $\beta_0=1.0$ and (c) $\beta_0=5.0$.  Figs. (d), (e), and (f) are the corresponding contour plots of (a), (b) and (c). The other parameters are same as in Fig.~\ref{fig1}.}
\label{fig5}
\end{figure*}
\begin{figure*}[!ht]
\begin{center}
\includegraphics[width=0.9\linewidth]{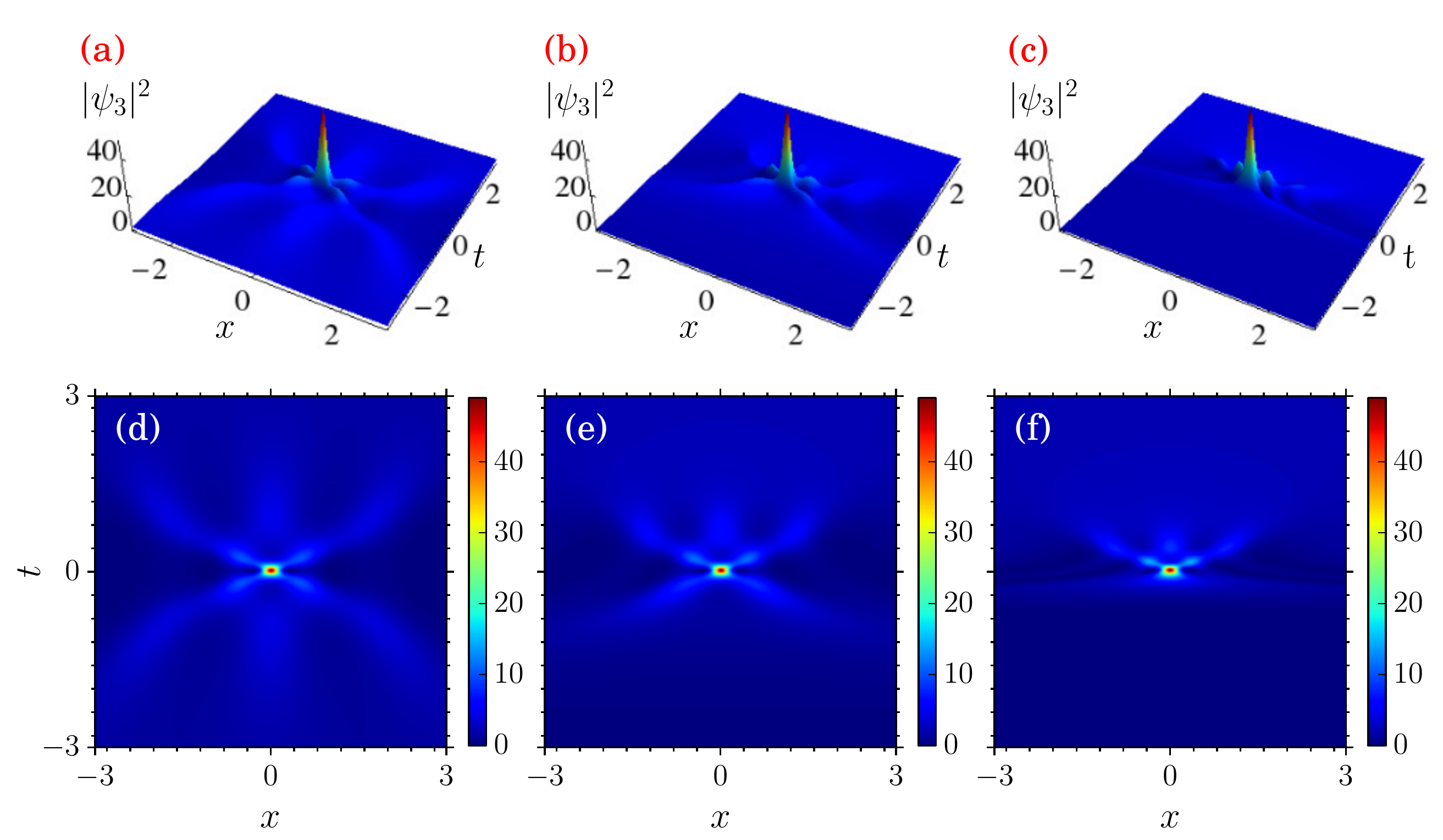}
\end{center}
\caption{(Color online) Third-order RWs for $R(t)=1+\tanh\big({\beta_0 t}/{2}\big)$ and $\beta(t)^2={\beta_0^2}/{2}\left[1-\tanh\left({\beta_0 t}/{2}\right)\right]$. The parameter $\beta_0$ is varied as (a) $\beta_0=0.1$, (b) $\beta_0=1.0$, and (c) $\beta_0=5.0$.  Figs. (d), (e), and (f) are their corresponding contour plots. The other parameters are same as in Fig.~\ref{fig1}.}
\label{fig6}
\end{figure*} 
For this choice, the relation (\ref{a5}) fixes the interatomic interaction term to be of the form $R(t)=1+\tanh \left({\beta_0 t}/{2}\right)$. The first, second, and third-order RW solutions for this trap frequency and strength of interatomic interaction are found to be 
\begin{align}
\label{a22}
\psi_j(x,t) = r_0\sqrt{1+\tanh\Big(\frac{\beta_0}{2}t\Big)}  \, U_j(X,T) \, \eta(x,t),
\end{align}%
where $j = 1, 2, 3$ and
\begin{widetext}
\begin{align}
\eta(x,t)  = \exp\left\{i \left[\frac{\beta_0\sech^2\left(\frac{\beta_0 t}{2}\right)x^2}{4\left[1+\tanh(\frac{\beta_0 t}{2})\right]}-c_1r_0^2\left[1+\tanh{\left(\frac{\beta_0 t}{2}\right)}\right]x +\frac{c_1^2r_0^4\left(\beta_0 t+2\log\left[\cosh(\frac{\beta_0 t}{2})-\tanh(\frac{\beta_0 t}{2})\right]\right)}{\beta_0}\right]\right\}. \notag
\end{align}
\end{widetext}
The qualitative nature of the first, second, and third-order RWs for $R(t)=1+\tanh\big({\beta_0 t}/{2}\big)$ and $\beta(t)^2={\beta_0^2}/{2}\left[1-\tanh\left({\beta_0 t}/{2}\right)\right]$ turns out to be the same as in the previous case (Fig.~\ref{fig8}(d), (h) and (l)) when the amplitude parameter $r_0$ is varied and so we do not display the outcome here separately.  On the other hand, we identify interesting structures while varying the parameter $\beta_0$ which is discussed in the following.

In Fig.~\ref{fig4}, we depict the first-order RW for these choices of $R(t)$ and $\beta(t)$.  When $\beta_0 = 0.1$ the first-order RW is as shown in Fig.~\ref{fig4}(a) (see also Fig.~\ref{fig4}(d) for the corresponding contour plot). By altering the value of the trap parameter $\beta_0$ to $1.0$, the structure of the first-order RW gets modified as shown in Fig.~\ref{fig4}(b)/Fig.~\ref{fig4}(e). One can also see from Fig.~\ref{fig4} that, as the trap parameter $\beta_0$ is increased, the RW gradually becomes more localized in time and the condensate atoms settle down to a slightly higher density background due to the attractive nature of the potential.  When $\beta_0 = 5.0$ the modified structure of first-order RW is given in Fig.~\ref{fig4}(c)/Fig.~\ref{fig4}(f), where this feature is even more prominent. The density profiles of the corresponding second-order RWs are presented in Fig.~\ref{fig5}.  When we tune the parameter $\beta_0$ from $0.1$ upwards the wave subcrests start to stretch.  From the contour plots we can observe that the stretches occur on one side of the RW only.  When we increase the value of $\beta_0$ further the second-order RW gets modified to a first-order RW like structure which is demonstrated in Fig.~\ref{fig5}(c).  A similar transition has also been observed in the third-order RW case as well which is illustrated in Fig.~\ref{fig6}.  The third-order RW acquires a new structure as shown in Fig.~\ref{fig6}(b)/Fig.~\ref{fig6}(e) when we increase the value of the parameter $\beta_0$ from $0.1$ to $1.0$.  At $\beta_0=5.0$, we observe that the third-order RW acquires a further modified structure which is displayed in Fig.~\ref{fig6}(c) and Fig.~\ref{fig6}(f).  Note the similarity in the central part with that of the first-order RW as given in Fig.~\ref{fig4}(c) and Fig.~\ref{fig4}(f).  

\subsection{Time-dependent periodic trap}
In the case of the third choice we consider the time-dependent periodic trap frequency to be of the form $\beta(t)^2=2\beta_0^2[1+3\tan^2(\beta_0 t)]$ so that the strength of the time-dependent periodic interatomic interaction turns out to be $R(t)=1+\cos{(2 \beta_0 t)}$. Substituting these two expressions in (\ref{a7}), we find
\begin{align}
\label{a24}
\psi_j(x,t)=r_0\sqrt{1+\cos{(2\beta_0 t)}}  \, U_j(X,T) \, \eta(x,t),
\end{align}
where 
\begin{align}
\eta(x,t)= & \exp\bigg\{i\bigg[\beta_0 \tan{(\beta_0 t)}x^2+2c_1 r_0^2  \cos{(\beta_0 t)^2} x \notag \\
& \left.\left.-\frac{c_1^2 r_0^4(12\beta_0 t+8\sin{(2\beta_0 t)}+\sin{(4\beta_0 t)})}{16\beta_0}\right]\right\}.\notag
\end{align}
Here $U_j(X,T)$, $ j = 1, 2, 3$, are again the first, second and third-order RW solutions of the NLS equation (vide Eqs. (\ref{a8}), (\ref{a11}), and (\ref{a12})).
\begin{figure*}[!ht]
\begin{center}
\includegraphics[width=0.8\linewidth]{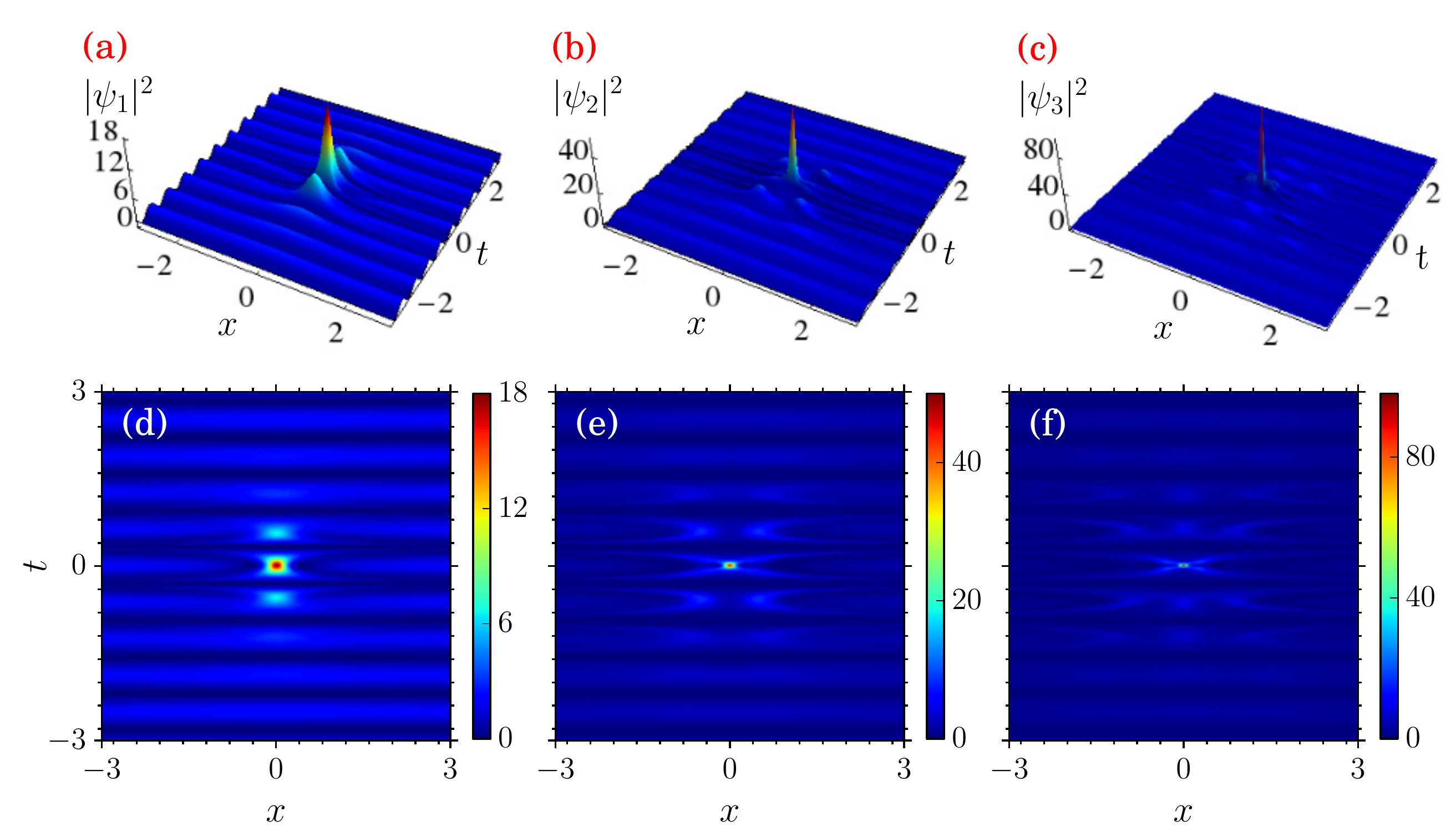}
\end{center}
\caption{(Color online) (a) First-order RW, (b) second-order RW, and (c) third-order RW for $R(t)=1+\cos{(2 \beta_0 t)}$ and $\beta(t)^2=2\beta_0^2[1+3\tan^2(\beta_0 t)]$.  (d), (e) and (f) are the corresponding contour plots of (a), (b) and (c). The parameters are $r_0=1.0$, $\beta_0=2.5$, $c_1=0.01$ and $\delta=0.01$.}
\label{fig7}
\end{figure*}

In Fig.~\ref{fig7}, we present the density profiles of the first, second, and third-order RWs (first row) and their corresponding contour plots (second row) for the strength of time-dependent interatomic interaction $R(t)=1+\cos{(2 \beta_0 t)}$ and time-dependent periodic trap frequency $\beta(t)^2=2\beta_0^2[1+3\tan^2(\beta_0 t)]$ .  Fixing the value of $\beta_0$ at $2.5$, from Fig.~\ref{fig7} we see that the RW exists on a periodic background for the above forms of $R(t)$ and $\beta^2(t)$.  Further we have also verified that the behaviour of the RWs as a function of $\beta_0$ follows the same qualitative picture discussed above for the other forms of $R(t)$ and $\beta^2(t)$.

\section{Characteristics of triplet RWs}

Very recently it has been shown that one can generalize the expressions for the higher order RW solutions of the scalar NLS equation (\ref{nls}) given in Appendix A further by introducing certain free parameters which allow one to split a symmetric form RW solution into a multi-peaked solution and that by varying these free parameters one can extract certain novel patterns of RWs \cite{ankie}.  The introduction of free parameters decompose the higher-order RW solutions into $n(n+1)/2$ first-order forms, where $n$ is the order of the RW, however maintaining the symmetry of the higher-order solutions even in their decomposed forms.  Further the free parameters are shown to determine the size and orientation of the first-order solutions \cite{ankie}. Triplets are symmetry preserving first-order structures revealing the fact that the (higher-) second-order RW solution is a family of three first-order rational solutions.  It is evident that the existence of triplets in an ocean corresponds to three big waves on the water surface in a row or `` three sister waves" \cite{juli}.  Inspired by such a possibility, in the following, we consider the second- and third-order RW solutions of (\ref{eq:1d-gp}) with suitable free parameters and analyze the symmetrical structures that arise due to these free parameters when we vary the strength of nonlinearity and the trap parameter. To begin with, we confine our attention to the second-order RW solution.  In this case, we have the following modified expressions \cite{ankie} for $G_2$, $H_2$ and $D_2$ in Eq. (\ref{a9}), that is
\begin{align}
 G_2 = &\,  12 \big[3-16 X^4-24 X^2(4 T^2+1)-48 l X-80 T^4 \notag \\
       &\, -72 T^2-48 m T\big], \notag \\
 H_2 = &\,  24 \big\{ T\big[ 15-16 X^4+24 X^2-48 l X-8(1-4 X^2)T^2  \notag \\  
       &\, -16 T^4 \big]+6m(1-4 T^2+4X^2)\big\}, \notag  \\
 D_2 = &\,  64 X^6+48 X^4 (4T^2+1)+12X^2(3-4T^2)^2+64 T^6\notag \\  
       &\, +432 T^4+396 T^2+9 +48m\big[18m+T(9+4 T^2 \notag \\ 
       &\, -12X^2)\big]+48l\big[(18l+X(3+12T^2 -4X^2)\big].  
\end{align}
Note that this RW solution now contains two free parameters, namely $l$ and $m$. The parameters $l$ and $m$ describe the relative positions of the first-order RWs in the triplet.  Substituting the above expressions in  (\ref{a15}), (\ref{a22}) and (\ref{a24}), for $j=2$, we obtain the corresponding second-order RW solutions to the GP equation (\ref{eq:1d-gp}) with the free parameters $l$ and $m$ included.  
\begin{figure*}[!ht]
\begin{center}
\includegraphics[width=0.8\linewidth]{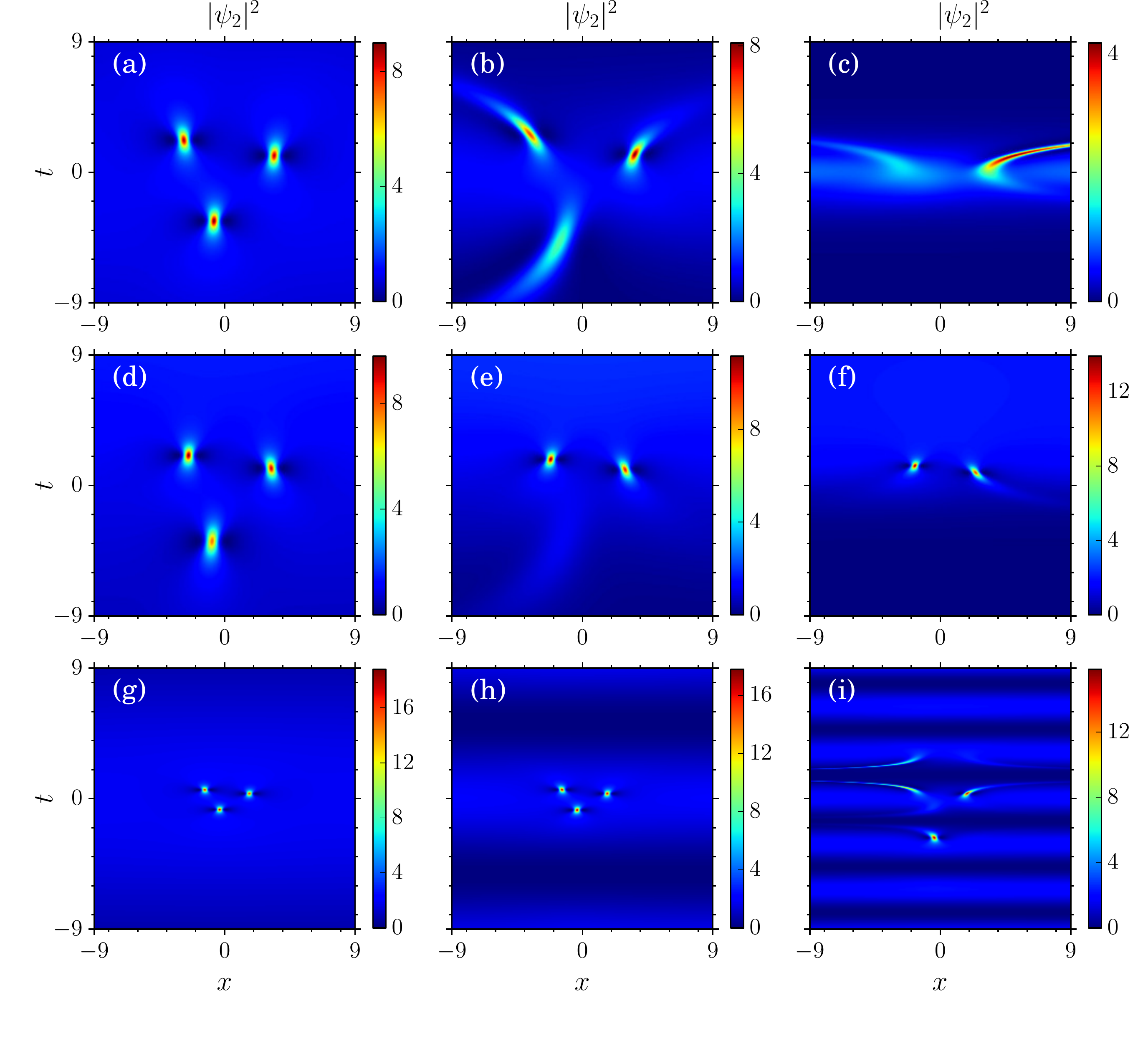}
\end{center}
\caption{(Color online) Triplet RWs. (a)-(c) $R(t)=\sech{(\beta_0 t+\delta)}$ and $\beta(t)^2=\beta_0^2$, (d)-(f) $R(t)=1+\tanh\big({\beta_0 t}/{2}\big)$ and $\beta(t)^2={\beta_0^2}/{2}\left[1-\tanh\left({\beta_0 t}/{2}\right)\right]$, (g)-(i) $R(t)=1+\cos{(2 \beta_0 t)}$ and $\beta(t)^2=2\beta_0^2[1+3\tan^2(\beta_0 t)]$.  Further, in (a), (d) and (g) $\beta_0=0.1$, in (b), (e) and (h) $\beta_0=0.3$ and in (c), (f) and (i) $\beta_0=1.0$.  The other parameters are $l=15$, $m=25$, $r_0=1.0$, $c_1=0.01$ and $\delta=0.01$.}
\label{fig8b}
\end{figure*}
When $l= m=0$, this solution coincides with the one given earlier (vide Eq. (\ref{a11})) which contains one largest crest and four subcrests with two deepest troughs (Fig. \ref{fig8}(h)).  When $l$ and $m$ are not equal to $0$, the second-order RW splits into three first-order RWs.  These waves emerge in a triangular fashion (a triplet pattern). The three first-order RWs form a triangular pattern with $120$ degrees of angular separation between them \cite{ankie}.  In Figs. \ref{fig8b}(a)-(c) we display the triplet pattern for $R(t)=\sech{(\beta_0 t+\delta)}$ and $\beta(t)^2=\beta_0^2$ when $l=15$ and $m=25$.  The triplet RW pattern for $\beta_0=0.1$ is shown in Fig.~\ref{fig8b}(a).  When we increase the parameter $\beta_0$ to $0.3$ we observe that the triplet pattern has started to collapse as shown in Fig.~\ref{fig8b}(b) and a complete collapse is observed as in Fig.~\ref{fig8b}(c) when we increase the $\beta_0$ value further to $1.0$.  Figs. \ref{fig8b}(d)-(g) represent the triplet pattern for $R(t)=1+\tanh\big({\beta_0 t}/{2}\big)$ and $\beta(t)^2={\beta_0^2}/{2}\left[1-\tanh\left({\beta_0 t}/{2}\right)\right]$.  The formation of triplet RWs is shown in Fig.~\ref{fig8b}(d) when $\beta_0=0.1$.  When we increase the value $\beta_0$ to $0.3$ one of the single RWs in the triplet pattern vanishes which is illustrated in Fig.~\ref{fig8b}(e).  By increasing the parameter $\beta_0$ further we observe that two first-order RWs are more localized in time as shown in Fig.~\ref{fig8b}(f).  Figs. \ref{fig8b}(g)-(i) represent the triplet pattern for $R(t)=1+\cos{(2 \beta_0 t)}$ and $\beta(t)^2=2\beta_0^2[1+3\tan^2(\beta_0 t)]$.  The form of the triplet pattern for $\beta_0=0.1$ is displayed in Fig.~\ref{fig8b}(g).  Here also when we increase the value of $\beta_0$ we observe the collapse of the triplet pattern in the periodic wave background (vide Fig.~\ref{fig8b}(i)).
\begin{figure*}[!ht]
\begin{center}
\includegraphics[width=0.9\linewidth]{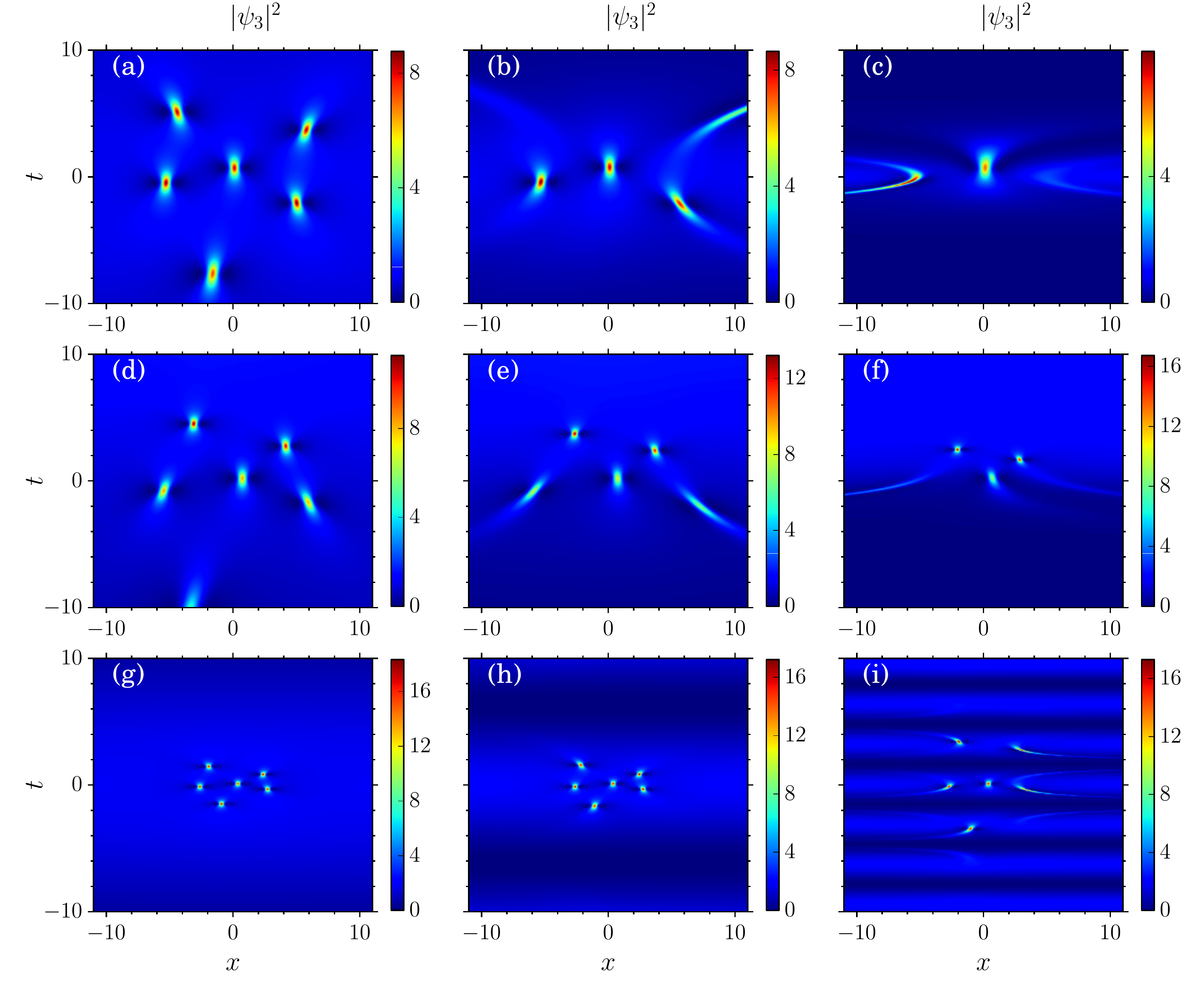}
\end{center}
\caption{(Color online) Sextet RWs. (a)-(c) $R(t)=\sech{(\beta_0 t+\delta)}$ and $\beta(t)^2=\beta_0^2$, (d)-(f) $R(t)=1+\tanh\big({\beta_0 t}/{2}\big)$ and $\beta(t)^2={\beta_0^2}/{2}\left[1-\tanh\left({\beta_0 t}/{2}\right)\right]$, (g)-(i) $R(t)=1+\cos{(2 \beta_0 t)}$ and $\beta(t)^2=2\beta_0^2[1+3\tan^2(\beta_0 t)]$.  Further, in (a), (d) and (g) $\beta_0=0.1$, in (b), (e) and (h) $\beta_0=0.25$ and in (c), (f) and (i) $\beta_0=1.0$.  The other parameters are $l=10$, $m=20$, $g=500$, $h=500$, $r_0=1.0$, $c_1=0.01$ and $\delta=0.01$.}
\label{fig8h}
\end{figure*}

We then move on to investigate the structure of the third-order RW solution with four free parameters, namely $l,m,g$ and $h$.  The third-order RW solution with four free parameters is much lengthier than the one without free parameters and so we do not give the explicit expression here and analyze the results only graphically.  Here also we analyze the solution with respect to the free parameters.  When $l= m= g= h=0$, we have the classical third-order RW solution which is shown in Fig.~\ref{fig8}(l).  It has one largest crest and six subcrests with two deepest troughs.  For non-zero values of $l$, $m$, $g$ and $h$, the third-order RW splits into six separated first-order RWs.  When we increase the value of the free parameters, the six first-order RWs take new positions.  The sextet pattern is displayed in Figs. \ref{fig8h}(a)-(c) for $R(t)=\sech{(\beta_0 t+\delta)}$ and $\beta(t)^2=\beta_0^2$ when $l=10$,  $m=20$, $g=500$ and $h=500$.  For $\beta_0=0.1$ the set of six first-order RWs is shown in Fig.~\ref{fig8h}(a).  When we increase the value of $\beta_0$ to $0.25$, three peaks disappear and the remaining peaks start to bend as shown in Fig.~\ref{fig8h}(b).  When we increase the value of $\beta_0$ further the RWs bend in the plane wave background as given in Fig. \ref{fig8h}(c). Figs. \ref{fig8h}(d)-(f) represent six first-order RWs for the time-dependent nonlinearity strength $R(t)=1+\tanh\big({\beta_0 t}/{2}\big)$ and the time-dependent external trap frequency $\beta(t)^2={\beta_0^2}/{2}\left[1-\tanh\left({\beta_0 t}/{2}\right)\right]$.  The formation of six first-order RWs at $\beta_0=0.1$ is shown in Fig.~\ref{fig8h}(d).  When we increase the strength of the parameter $\beta_0$ to $0.25$, one of six first-order RWs vanishes as seen in Fig.~\ref{fig8h}(e).  If we increase the value $\beta_0$ further three out of six peaks bend in the plane wave background and eventually collapse which is shown in Fig.~\ref{fig8h}(f).  Figs.~\ref{fig8h}(g)-(i) represent the sextet pattern for $R(t)=1+\cos{(2 \beta_0 t)}$ and $\beta(t)^2=2\beta_0^2[1+3\tan^2(\beta_0 t)]$.  The form of the six first-order RWs for $\beta_0=0.05$ is displayed in Fig.~\ref{fig8h}(g) and for further increase in $\beta_0$, the modified structures in the periodic wave background are as shown in Figs. \ref{fig8h}(h) and \ref{fig8h}(i).

\section{Characteristics of Breathers}

In the previous two sections we have analyzed how the RW profiles get modified by the variations of the distributed coefficients present in the  variable coefficient NLS Eq. (\ref{eq:1d-gp}).  In this section we analyze how the breather structures get modified in the condensates when we vary the strength of the external trap parameter. 

To begin with we consider the first-order breather solution of the NLS equation (\ref{nls}), which is given in \cite{eleon} and is a special case of the GB solution (\ref{nls2}), 
\begin{align}
\tilde{U_1}(X,T) = & \,\biggr[\frac{f^2 \cosh[\alpha (T - T_1)] + 2 i \,f v \sinh[\alpha (T - T_1)]}{2  \cosh[\alpha (T - T_1)] - 2v \cos[ f (X - X_1) ]} \notag \\
& \, -1\biggr] \times \exp{(iT)},
\label{b1}
\end{align}%
where the parameters $f$ and $v$ are expressed in terms of a complex eigenvalue (say $\lambda$), that is $f=2\sqrt{1+\lambda^2}$ and $v = \mbox{Im}(\lambda)$, and $X_1$ and $T_1$ serve as coordinate shifts from the origin.  The real part of the eigenvalue represents the angle that the one-dimensionally localized solutions form with the $T$ axis, and the imaginary part characterizes frequency of periodic modulation.  The parameter $\alpha$ $(=f v)$ in (\ref{b1}) is the growth rate of modulation instability.  Substituting this breather solution of the NLS equation into (\ref{a15}), (\ref{a22}) and (\ref{a24}), we study the underlying dynamics of (\ref{eq:1d-gp}).   
\begin{figure}[!ht]
\begin{center}
\includegraphics[width=0.95\linewidth]{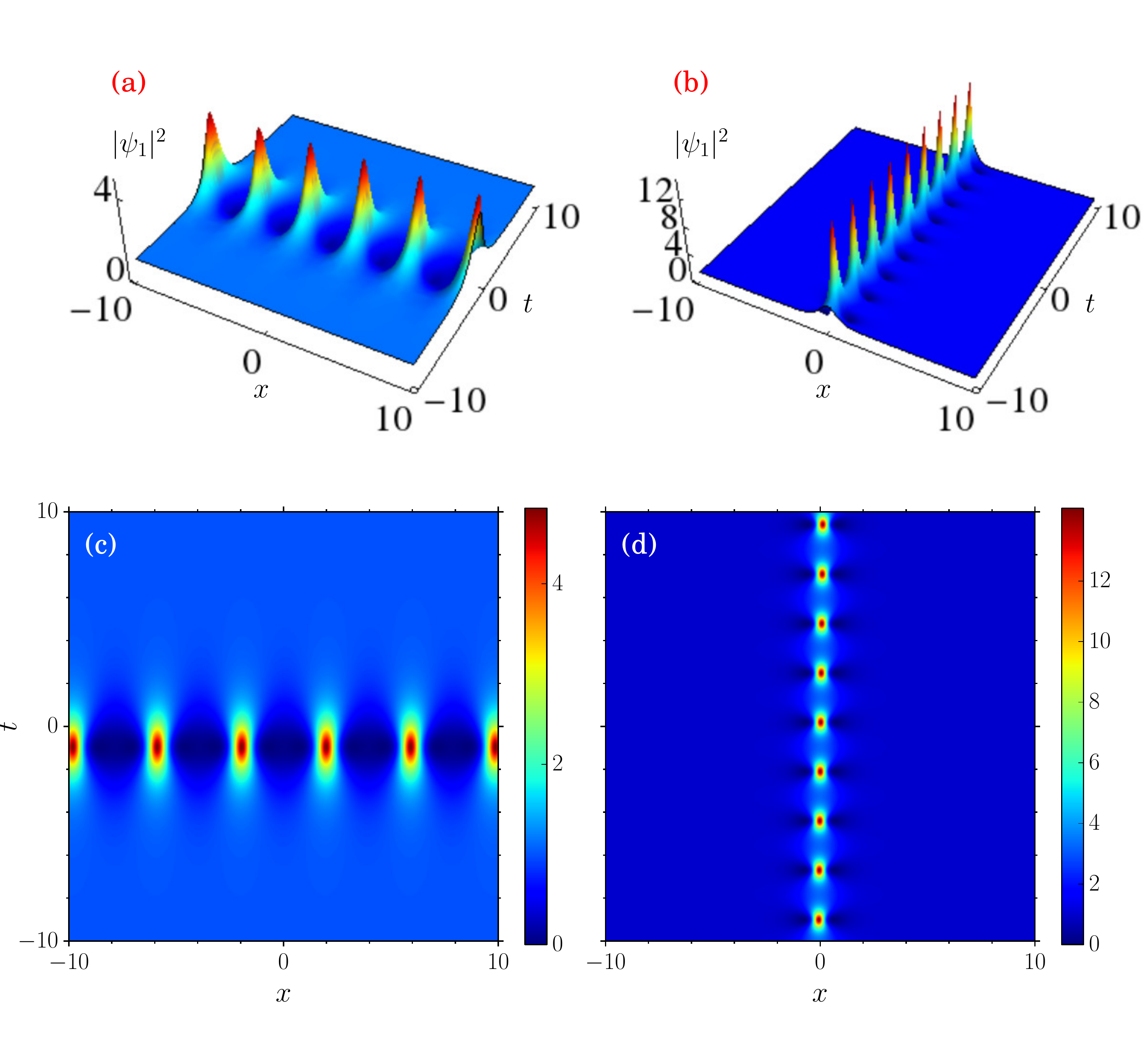}
\end{center}
\caption{(Color online) (a) AB for $\lambda=0.6 i$, (b) MB for $\lambda=1.4 i$ when $R(t)=\sech{(\beta_0 t+\delta)}$ and $\beta(t)^2=\beta_0^2$. (c) and (d) are their corresponding contour plots. The other parameters are $r_0=1.0$, $\beta_0=0.01$, $c_1=0.01$, and $\delta=0.01$. }
\label{fig9}
\end{figure}

\begin{figure}[!ht]
\begin{center}
\includegraphics[width=0.95\linewidth]{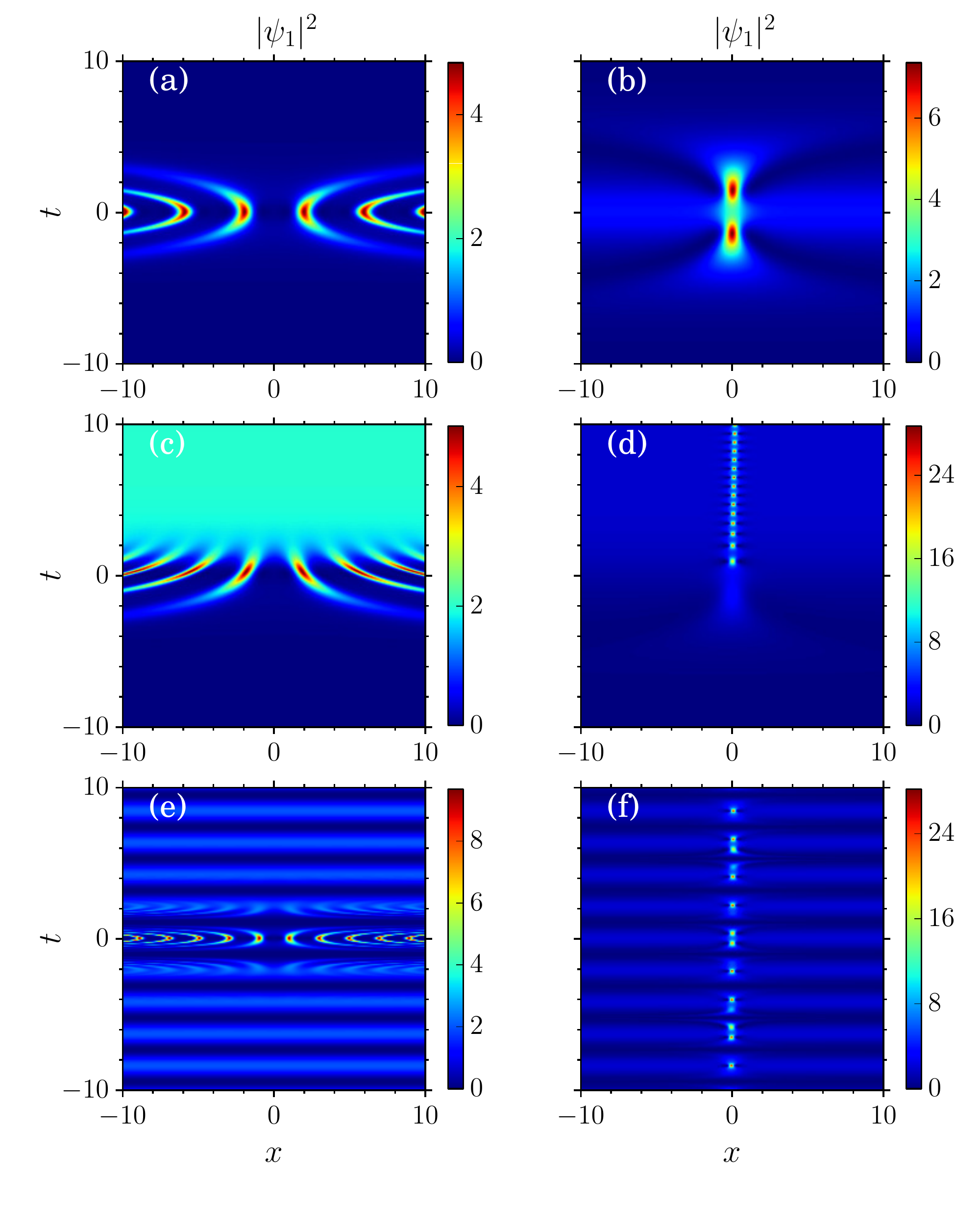}
\end{center}
\caption{(Color online) Stretching of breathers.  (a), (c) and (e) AB with an eigenvalue $\lambda=0.6 i$, (b), (d) and (f) MB with $\lambda=1.4 i$ for the three different forms of $R(t)$ discussed in the text.  (a)-(d) $\beta_0=0.8$, (e)-(f) $\beta_0=1.5$.  The other parameters are $r_0=1.0$, $c_1=0.01$ and $\delta=0.01$.}
\label{fig9a}
\end{figure}

For illustration, we consider the case $R(t)=\sech{(\beta_0 t+\delta)}$ and $\beta(t)^2=\beta_0^2$, and plot the outcome in Fig.~\ref{fig9}.  The first row in Fig.~\ref{fig9} represents an Akhmediev breather (AB) solution for the eigenvalue $\lambda=0.6 i$ and Ma breather (MB) solution for $\lambda=1.4 i$ and the second row represents their contour plots.  The AB and MB solutions are localized in time and space, respectively, as discussed in Sec. I.  This is clearly demonstrated in the two columns of Fig.~\ref{fig9}.  When we tune the parameter $\beta_0$, a new structure against a breather background is obtained.  To visualize this we fix the value of $\beta_0$ to be $0.8$.  For this value a stretching occurs in space in the case of AB (Fig.~\ref{fig9a}(a)) whereas in the case of MB (Fig.~\ref{fig9a}(b)) the stretching occurs in time.  Figs. \ref{fig9a}(c)-(d) illustrate the AB and MB profiles for the time-dependent nonlinearity coefficient $R(t)=1+\tanh\big({\beta_0 t}/{2}\big)$ and time-dependent trap frequency $\beta(t)^2=({\beta_0^2}/{2})\left[1-\tanh\left({\beta_0 t}/{2}\right)\right]$.  Here also we tune the strength of the trap parameter $\beta_0$ to 0.8 and observe that stretching occurs over space in AB which is depicted in Fig.~\ref{fig9a}(c) and the MB gets more localized and when $t\leq 0$ the breather profile completely disappears as shown in Fig.~\ref{fig9a}(d).  Our results reveal the fact that when we tune the parameter $\beta_0$ in the obtained breather solution, the breather gets a modified structure corresponding to a distortion of the breather profile. 
Figs. \ref{fig9a}(e) and (f) respectively represent the AB and MB in the periodic background for $R(t)=1+\cos{(2 \beta_0 t)}$ and $\beta(t)^2=2\beta_0^2[1+3\tan^2(\beta_0 t)]$ when $\beta_0=1.5$.  

\begin{figure*}[!ht]
\begin{center}
\includegraphics[width=0.8\linewidth]{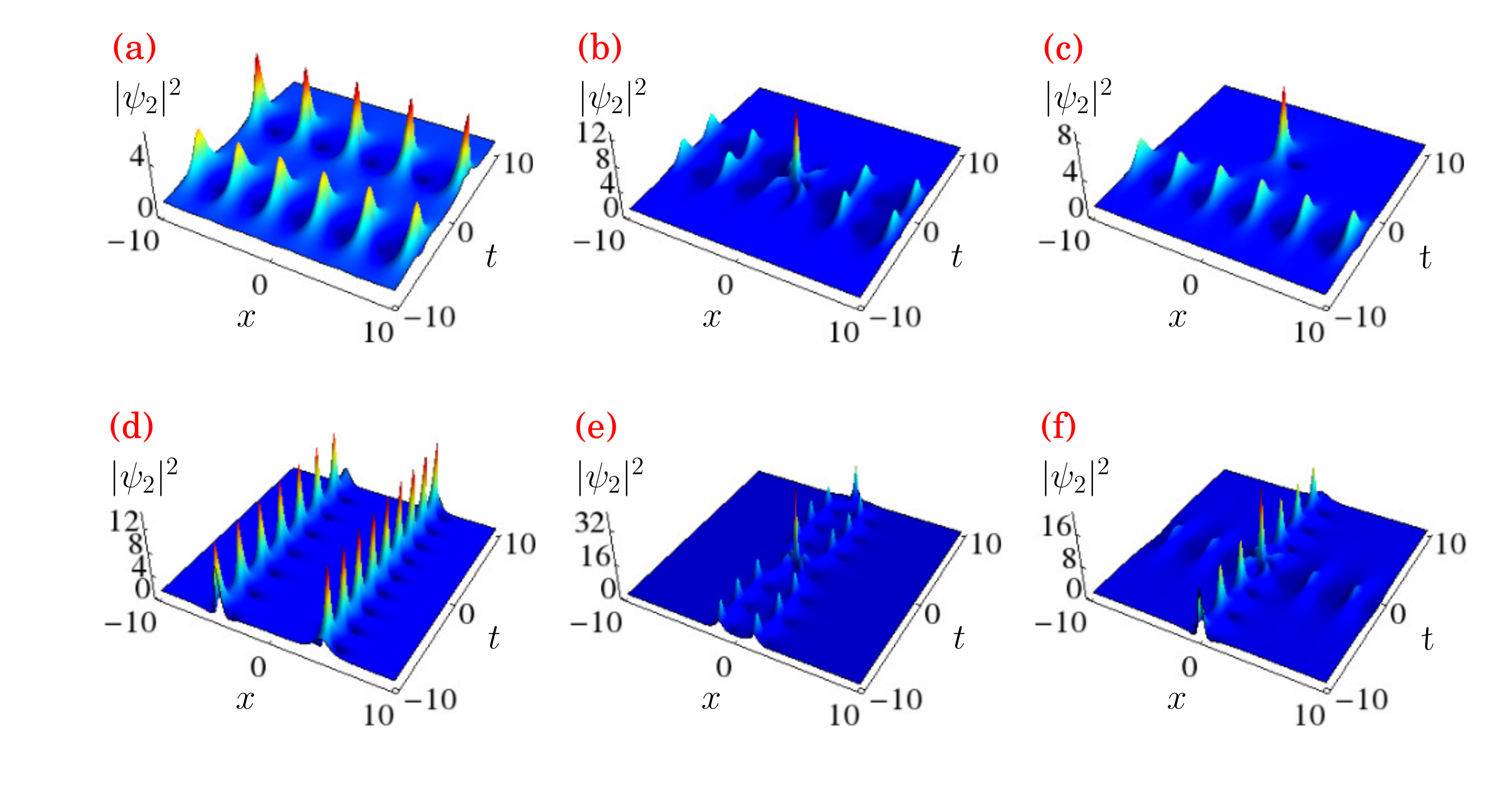}
\end{center}
\caption{(Color online) (a) Two AB profiles for $\lambda_1=0.55i$ and $\lambda_2=0.75i$ with $T_1=-3$ and $T_2=3$, (b) Two ABs without time shifts, (c) AB-RW profile for $\lambda_1=0.55i$ and $\lambda_2=0.99i$ (d) Two MBs for $\lambda_1=1.3i$ and $\lambda_2=1.4i$ with $X_1=-3$ and $X_2=3$, (e) Two MBs without space shifts and (f) the intersection of AB and MB for $\lambda_1=0.5i$ and $\lambda_2=1.3i$ for $R(t)=\sech{(\beta_0 t+\delta)}$ and $\beta(t)^2=\beta_0^2$. The other parameters are same as in Fig.~\ref{fig1}.}
\label{br2}
\end{figure*}

Next, we proceed to construct the two-breather solutions of (\ref{eq:1d-gp}) and analyze how these solutions are distorted by the variations of modulation parameters.  The two-breather solution of the NLS equation is given by \cite{kadz}
\begin{align}
\label{2b1}
\tilde{U_2}(X,T)=\left[1+\frac{\tilde{G_2}(X,T)+i \tilde{H_2}(X,T)}{\tilde{D_2}(X,T)}\right]\exp{(iT)},
\end{align}
where $\tilde{G_2}$, $\tilde{H_2}$, and $\tilde{D_2}$ are given by 
\begin{subequations}
\begin{align}
\tilde{G_2} = &\, -(k_1^2-k_2^2) \biggr[\frac{k_1^2\delta_2}{k_2}\cosh(\delta_1T_{s1})\cos(k_2X_{s2}) \notag \\ 
              &\,  - (k_1^2-k_2^2)\cosh(\delta_1T_{s1})\cosh(\delta_2T_{s2}) \notag \\ 
              &\, -\frac{k_2^2\delta_1}{k_1}\cosh(\delta_2T_{s2})\cos(k_1 X_{s1})\biggr],  
\end{align}
\begin{align}
\tilde{H_2} = &\, -2(k_1^2-k_2^2)\biggr[\frac{\delta_1\delta_2}{k_2}\sinh(\delta_1 T_{s1})\cos(k_2 X_{s2})\notag \\ 
              &\, - \frac{\delta_1 \delta_2}{k_1}\sinh(\delta_2T_{s2}) \cos(k_1 X_{s1}) \notag \\   
              &\,  -\delta_1 \sinh(\delta_1 T_{s1})\cosh(\delta_2 T_{s2}) \notag \\   
              &\,  + \delta_2 \sinh(\delta_2 T_{s2})\cosh(\delta_1 T_{s1})\biggr], 
\end{align}
\begin{align}
\tilde{D_2} = &\,  2 (k_1^2 + k_2^2) \frac{\delta_1 \delta_2}{k_1 k_2} \cos(k_1X_{s1})\cos(k_2 X_{s2}) \notag \\ 
              &\,  + 4 \delta_1 \delta_2 (\sin(k_1 X_{s1}) \sin(k_2 X_{s2})  \notag \\ 
              &\,  + \sinh(\delta_1 T_{s1})\sinh(\delta_2 T_{s2})) \notag \\ 
              &\,- (2 k_1^2 - k_1^2 k_2^2 + 2 k_2^2) \cosh(\delta_1 T_{s1})\cosh(\delta_2 T_{s2}) \notag \\ 
              &\, - 2 (k_1^2 - k_2^2) \biggr[\frac{\delta_1}{k_1}\cos(k_1 X_{s1})\cosh(\delta_2 T_{s2}) \notag \\ 
              &\,  - \frac{\delta_2}{k_2}\cos(k_2 X_{s2}) \cosh(\delta_1 T_{s1})\biggr], 
\end{align}
\end{subequations}
where the modulation frequencies, $k_j=2\sqrt{1+\lambda_j^2}$, $j=1,2$, are described by the (imaginary) eigenvalues $\lambda_j$.  In the above expressions, $X_j$, $T_j$, $j=1,2$, represent the shifted point of origin, $\delta_j$ $(=k_j\sqrt{4-k_j^2}/2)$ is the instability growth rate of each component and $X_{sj}=X-X_j$ and $T_{sj}=T-T_j$ are shifted variables.     
\begin{figure*}[!ht]
\begin{center}
\includegraphics[width=0.8\linewidth]{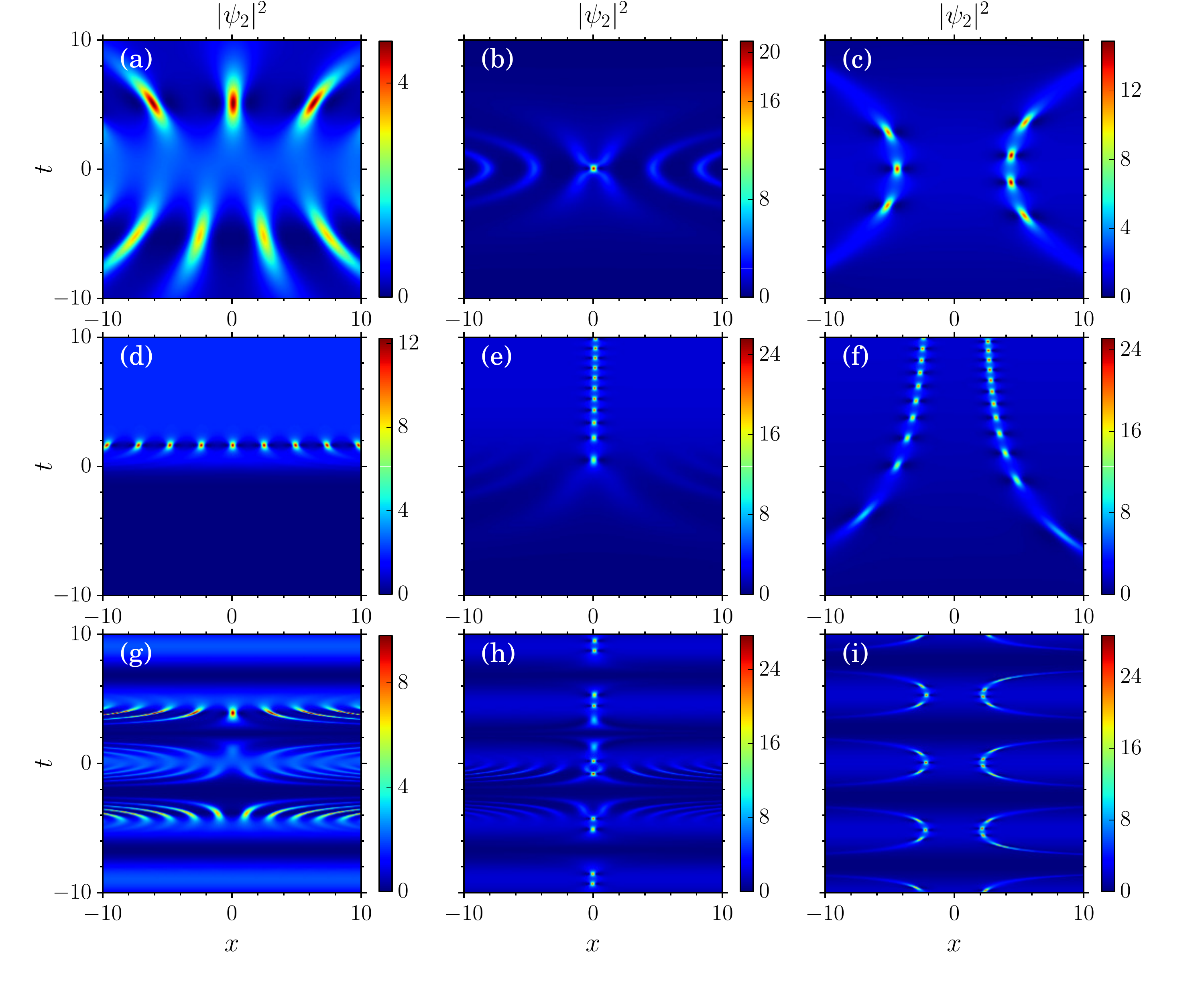}
\end{center}
\caption{(Color online) Distortion of two-breather profiles.  (a), (d) and (g) two ABs with $\lambda_1=0.55 i$ and $\lambda_2=0.75 i$, (b), (e) and (h) the intersection of AB-MB profiles with $\lambda_1=0.5 i$ and $\lambda_2=1.3 i$ and (c), (f) and (i) two MBs with $\lambda_1=1.3 i$ and $\lambda_2=1.4 i$.  The trap parameter $\beta_0$ is chosen as (a) $0.15$, (b) and (e) $0.5$, (c) and (f) $0.2$, (d) $1.5$ and (g)-(i) $0.7$.  The other parameters are $r_0=1.0$, $c_1=0.01$ and $\delta=0.01$.}
\label{br2a}
\end{figure*}

With two purely imaginary eigenvalues, $\lambda_j$, $j=1,2$, the solution (\ref{2b1}) is capable of describing a variety of possible second-order breather structures.  The solution includes ABs, MBs and the intersection of AB and MB solutions for certain combinations of eigenvalues.  For example, when the imaginary parts of both  the eigenvalues $\mbox{Im}(\lambda_j)$, $j=1,2$, lie between $0$ and $1$, we obtain the ABs. On the other hand when both of them are greater than one ($\mbox{Im}(\lambda_j)>1$) we obtain the MBs and in the mixed possibility case, that is one of the eigenvalues is less than one ($\mbox{Im}(\lambda_1)<1$) and the other eigenvalue ($\mbox{Im}(\lambda_2>1$) is greater than one, we obtain the intersection of AB and MB solutions.  

Substituting the two-breather NLS solution (\ref{2b1}) in (\ref{a15}), (\ref{a22}) and (\ref{a24}) we obtain the general two-breather solution of (\ref{eq:1d-gp}).  Fig.~\ref{br2} displays the evolution of the two-breather solution of (\ref{eq:1d-gp}) for $R(t)=\sech{(\beta_0 t+\delta)}$ and $\beta(t)^2 =\beta_0^2 = (0.01)^2 $ with imaginary eigenvalues.  To obtain the ABs from (\ref{2b1}) we consider the situation where the magnitudes of both the eigenvalues $\lambda_1$ and $\lambda_2$ are less than $1$ ($\lambda_1=0.55i$ and $\lambda_2=0.75i$).  One AB developing with a time delay after another is shown in Fig.~\ref{br2}(a), while in Fig.~\ref{br2}(b) we present the case where there is no such time delay.  In Fig. ~\ref{br2}(c) we depict the case when one AB along with a RW coexist for the choice $\lambda_1=0.55i$ and $\lambda_2=0.99i$.  When we take the eigenvalues $\lambda_1$ and $\lambda_2$ to be $1.3i$ and $1.4i$, respectively, we obtain two MB solutions.  Similarly the evolution of the two MBs with and without spatial delay is shown in Figs. \ref{br2}(d) and \ref{br2}(e), respectively.  We also observe that the distance between the MBs increases when we set both the eigenvalues to be nearly equal, say for example $\lambda_1=1.3i$ and $\lambda_2=1.31i$ which is not displayed here.  When we take the eigenvalues as $\lambda_1=0.5i$ and $\lambda_2=1.3i$, the AB intersects with the MB which is displayed in Fig.~\ref{br2}(f).  When we tune the strength of the trap parameter $\beta_0$ to $0.15$, both the ABs get stretched in the plane wave background which is demonstrated in Fig.~\ref{br2a}(a).  When $\beta_0=0.5$, in the intersection of AB and MB solutions we observe that the AB gets a bending structure while the MB fully disappears in the plane wave background which is shown in Fig.~\ref{br2a}(b).  In Fig.~\ref{br2a}(c) we note that both the MBs develop a bending structure in the plane wave background when $\beta_0=0.2$.  Figs. \ref{br2a}(d)-(f) display the evolution of the two-breather solution of (\ref{eq:1d-gp}) for $R(t)=1+\tanh\big({\beta_0 t}/{2}\big)$ and $\beta(t)^2={\beta_0^2}/{2}\left[1-\tanh\left({\beta_0 t}/{2}\right)\right]$ with the imaginary eigenvalues.  When $\beta_0 = 0.15$ one of the ABs gets stretched which is not shown here and on further increase of the value of $\beta_0$ to $1.5$, one of the ABs gets annihilated and the other AB bends in the plane wave background which is demonstrated in Fig.~\ref{br2a}(d).  When $\beta_0 = 0.5$ the intersection of AB-MB structures is as shown in Fig.~\ref{br2a}(e).  In Fig. ~\ref{br2a}(f) we observe that both the MBs get a bending structure in the plane wave background.  Figs. \ref{br2a}(g)-(i) show the evolution of the two-breather solution of (\ref{eq:1d-gp}) for $R(t)=1+\cos{(2 \beta_0 t)}$ and $\beta(t)^2=2\beta_0^2[1+3\tan^2(\beta_0 t)]$ with the imaginary eigenvalues.  Here also when we increase the value of $\beta_0$ we observe the collapse of the two-breather solution in the periodic wave background.

\section{Conclusion}

In this work, we have constructed higher order RW solutions with and without free parameters for the quasi one-dimensional GP equation with time-dependent interatomic interation and external trap through the similarity transformation technique. By mapping the variable coefficient NLS equation onto the constant coefficient NLS equation we have derived these solutions.  We have shown that the mapping can be done when the external trap and the nonlinearly interatomic interaction of atoms satisfy a constraint.  From the known higher order RW and breather solutions of the constant coefficient NLS equation, we have derived the solutions of (\ref{eq:1d-gp}).  In our analysis, we have considered the harmonic trap frequency in three different forms, namely (i) time-independent expulsive trap, (ii) time-dependent monotonous trap, and (iii) time-dependent periodic trap, and correspondingly fixed the effective scattering length.  We then studied the characteristics of the constructed RW solutions in detail.  We have observed that the second- and third-order RWs transform to first-order RW-like structures when a parameter appearing in the harmonic trap (time-independent and time-dependent traps) is varied.  We have then analyzed the characteristics of triplet and sextet patterns of matter RWs for (\ref{eq:1d-gp}).  We have also constructed one-breather and two breather solutions of (\ref{eq:1d-gp}).  We have investigated how these periodic localized waves change in the plane wave background when we tune the trap parameter in the obtained breather solutions.  Our results may provide possibilities to manipulate RWs experimentally in a BEC system.

\acknowledgments

KM thanks the University Grants Commission (UGC-RFSMS), Government of India, for providing a research fellowship.  The work of PM forms part of Department of Science and Technology (Ref. No. SR/S2/HEP-03/2009) and Council of Scientific and Industrial Research (Ref. No. 03(1186)/10/EMR-II), Government of India funded research projects.  The work of MS forms part of a research project sponsored by NBHM, Government of India. The work forms part of an IRHPA project and a Ramanna Fellowship project of ML, sponsored by the Department of Science and Technology (DST), Government of India, who is also supported by a DAE Raja Ramanna Fellowship.

\begin{appendix}
\section{}
In the absence of the trap, and the nonlinearity strength $R(t)$ is equal to one, Eq. (\ref{eq:1d-gp}) reduces to the standard NLS equation $(\ref{nls})$.  Several localized and periodic structures of standard NLS are documented in the literature \cite{akmv:anki,eleon,kadz}.  Equation $(\ref{nls})$ admits $N^{th}$  order RW solution.  We present the RW solution of the NLS equation in the following form \cite{akmv:anki},
\begin{align}
\label{a6}
U_j(X,T)=\left[(-1)^j+\frac{G_j(X,T)+iTH_j(X,T)}{D_j(X,T)}\right]\exp{(iT)},
\end{align}
where $j=1,2,..,G_j,H_j$ and $D_j$ are polynomials in the variables $T$ and $X$.

\begin{figure}[!ht]
\begin{center}
\includegraphics[width=0.99\linewidth]{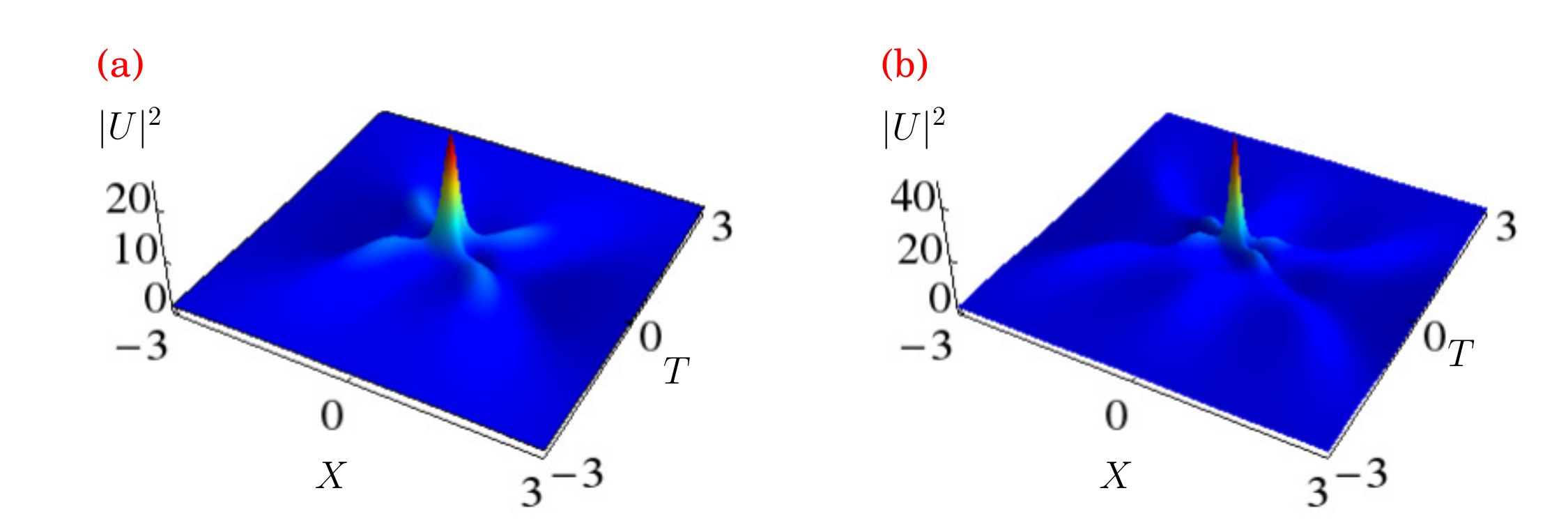}
\end{center}
\caption{(Color online) Profiles. (a) Second-order RW and (b) Third-order RW of the standard NLS equation.}
\label{nls-rog}
\end{figure}

For the first-order $(j=1)$ RW solution $G_1=4$, $H_1=8$ and $D_1=1+4X^2+4T^2$.  From  (\ref{a6}) we get $U_1=\left(4\frac{1+2iT}{1+4X^2+4T^2}-1\right)\exp{(iT)}$.  For covenience we multiply this expression by $-1 = \exp[i\pi]$ and consider the solution in the form
\begin{align}
\label{a8}
U_1(X,T)=\left(1-4\frac{1+2iT}{1+4X^2+4T^2}\right)\exp{[iT]}.
\end{align}
We use only this form of expression in our analysis. This solution is same as the one given in Eq. (\ref{nls3}) when $k=0$, $\omega=1$ and $\rho_0=1$.  
For the second-order $(j=2)$ RW solution, the function $G_2$, $H_2$, and $D_2$ turn out to be \cite{akmv:anki}
\begin{align}
\label{a9}
G_2 = & \frac{3}{8}-3X^2-2X^4-9T^2-10T^4-12X^2T^2, \notag \\
H_2 = & \frac{15}{4}+6X^2-4X^4-2T^2-4T^4-8X^2T^2
\end{align}
and
\begin{align}
D_2 = &  \frac{1}{8}\left(\frac{3}{4}+9X^2+4X^4+\frac{16}{3}X^6+33T^2+36T^4 \right. \notag \\
& \left.+\frac{16}{3}T^6-24X^2T^2+16X^4T^2+16X^2T^4\right). \notag
\end{align}
Then we get
\begin{align}
\label{a11}
U_2(X,T)=\left[1+\frac{G_2+iTH_2}{D_2}\right]\exp{(iT)}.
\end{align}
The profiles of second-order RW and third-order RW of the constant coefficient NLS equation are shown in Fig. \ref{nls-rog}. 
For the third-order $(j=3)$ RW solution, we have
\begin{align}
\label{a12}
U_3(X,T)=\left[-1+\frac{G_3+iTH_3}{D_3}\right]\exp{(iT)},
\end{align}
where
\begin{align}
G_3(X,T)= & \, g_0 + (2 T)^2 g_2 + (2 T)^4 g_4 + (2 T)^6 g_6 \notag \\ 
          & + (2 T)^8 g_8  + (2 T)^{10} g_{10},
\end{align}
with
\begin{align}
\label{solg}
g_0 = &  1 - (2 X)^2 - \frac{2}{3} (2 X)^4 + \frac{14}{45} (2 X)^6 + \frac{(2 X)^8}{45} + \frac{(2 X)^{10}}{675},\notag \\
g_2 = &  -3 - 20 (2 X)^2 + \frac{2}{3}(2 X)^4 - \frac{4}{45} (2 X)^6 + \frac{(2 X)^8}{45},\notag \\
g_4 = &  2 \left[-\frac{17}{3} + 5 (2 X)^2 - \frac{(2 X)^4}{3^2} + \frac{(2 X)^6}{3^3}\right]\notag \\
g_6 = &  \frac{2}{45}\left[73 +14 (2 X)^2 + \frac{7}{3}(2 X)^4\right],\notag \\
g_8 = &  \frac{1}{15} (11 + (2 X)^2), \;\;\;\; 
g_{10} = \frac{11}{675}
\end{align}
and 
\begin{align}
H_3(X,T)= & \, h_0 + (2 T)^2 h_2 + (2 T)^4 h_4 + (2 T)^6 h_6 \notag \\ 
          & + (2 T)^8 h_8  + (2 T)^{10} h_{10}, 
\end{align}
with
\begin{align}
h_0 = & 2 \left[7 + 7 (2 X)^2 - 2 (2 X)^4 - \frac{2}{3^2} (2 X)^6 - \frac{(2 X)^8}{45} \right. \notag \\
      &  \left. + \frac{(2 X)^{10}}{675} \right], \notag \\
h_2 = &  \frac{2}{3} \left[-11 - 28 (2 X)^2 - 2 (2 X)^4 - \frac{28}{45} (2 X)^6 + \frac{(2 X)^8}{45}\right],\notag \\
h_4 = &  \frac{4}{15} \left[-107 + 19 (2 X)^2 - \frac{7}{3} (2 X)^4 + \frac{(2 X)^6}{3^2}\right],\notag \\
h_6 = &  \frac{4}{45} \left[-29 - 2 (2 X)^2 +\frac{(2 X)^4}{3}\right],\notag  
\end{align}
\begin{align}
h_8 = & \frac{2}{3^3} \left[1 + \frac{(2 X)^2}{5}\right], \;\;\; h_{10} =\frac{2}{675}.
\end{align}
The denominator is represented by
\begin{align}
D_3(X,T)=& \, d_0 + (2 T)^2 d_2 + (2 T)^4 d_4 + (2 T)^6 d_6 + (2 T)^8 d_8 \notag \\ 
          &  + (2 T)^{10} d_{10}+ (2 T)^{12} d_{12},\notag \\
\end{align}
where
\begin{align}
d_0 = & \frac{1}{2^3} \left[1 + 6 (2 X)^2 + \frac{5}{3} (2 X)^4 + \frac{52}{45} (2 X)^6 + \frac{(2 X)^8}{15} \right.\notag \\ &  \left. + \frac{2}{675} (2 X)^{10} + \frac{(2 X)^{12}}{2025}\right],\notag 
\end{align}
\begin{align}
d_2 = &  23 - 9 (2 X)^2 + \frac{10}{3} (2 X)^4 + \frac{2}{15} (2 X)^6 - \frac{(2 X)^8}{45} \notag \\
      & + \frac{(2 X)^{10}}{675},\notag \\
d_4 = &  2 \left[71 + \frac{116}{3} (2 X)^2 - \frac{2}{3} (2 X)^4 - \frac{4}{45} (2 X)^6 + \frac{(2 X)^8}{135}\right],\notag 
\end{align}
\begin{align}
d_6 = &  \frac{32}{3} \left[\frac{17}{3} + 5 (2 X)^2 +\frac{(2 X)^4}{45} + \frac{(2 X)^6}{135}\right],\notag \\
d_8 = &  \frac{32}{15} \left[\frac{83}{3} + 2 (2 X)^2 +\frac{(2 X)^4}{3^2}\right),\notag \\
d_{10} = &  \frac{2^8}{225} \left[7 + \frac{(2 X)^2}{3}\right], \;\;\; d_{12} = \frac{2^9}{2025}.
\end{align}
\end{appendix}

\end{document}